\documentclass[reqno,aps,prd,preprint,showpacs,showkeys,preprintnumbers,amsmath,amssymb,amsfonts]{revtex4}
\newcommand{\wh}{\widehat}
\usepackage{graphics}
\usepackage[dvips]{graphicx}
\usepackage{dcolumn}
\usepackage{bm}
\usepackage[bookmarksnumbered]{hyperref}
\usepackage{color}

\numberwithin{equation}{section}

\begin{document}

\preprint{PRF1.30}

\pdfbookmark{Periodic relativity: the theory of gravity in flat space time}{tit}
\title{Periodic relativity: the theory of gravity in flat space time}


\author{Vikram H. Zaveri}
\homepage[]{http://orcid.org/0000-0002-4025-4721}

\email{zaverivik@hotmail.com}

\affiliation{Independent Researcher, B-4/6, Avanti Apt., Harbanslal Marg, Sion, Mumbai 400022 INDIA}


\date{January 11, 2025}

\begin{abstract}
In periodic relativity (PR), the curved space time of general relativity are eliminated by making use of a flat metric without weak field approximation. PR satisfies Einstein's field equations. Theory allows every two body system to deviate differently from the flat Minkowski metric. PR proposes a definite connection between the proper time interval of an object and Doppler frequency shift of its constituent particles as the object makes a relative motion with respect to the rest frame of the coordinate time. This is because fundamentally time is periodic in nature. Coordinate and proper time in GR are linear time. Periodic time of PR is the key parameter in development of quantum gravity theory in which the universe begins with a quantum fluctuation in the fundamental substance of the universe which is infinite, motionless and indivisible. PR is based on the dynamic weak equivalence principle which equates the gravitational mass with the relativistic mass. PR provides accurate solutions for the rotation curves of galaxies and the energy levels of the  Hydrogen spectra including Lamb shift using common formalism. Flat space time with Lorentz invariant acceleration presented here makes it possible to unite PR with quantum mechanics. PR satisfies Einstein's field equations with respect to the three major GR tests within the solar system and with respect to the derivation of Friedmann equation in cosmology. PR predicts limiting radius of the event horizon of a black hole to be $1R_g$ and the shadow of a black hole to be due to frequency shift of visible light into ultraviolet and higher range. PR equations can probe inside the event horizon. Mathematical proof of periodic nature of time is presented by way of introducing gravity into the electromagnetic wave formalism. Theory shows that the electromagnetic wave is held together by gravitational forces. Gravity assisted EM wave stacking creates $\gamma$-rays in M87 BH. Theory explains the mechanism of gravitational redshift predicted by general relativity and predicts powerful gravitational radiation at Planck epoch. Gravitational-wave strain amplitude is derived using quantum mechanical formalism. Bound on graviton mass is $m_g<1.51\times 10^{-41}$eV/$c^2$. Mechanism that explains Gamma-ray burst associated with GW150914 is presented. GRB can shrink the event horizon. 
\end{abstract}

\pacs{$04.50.Kd$,\: $04.20.Cv$,\: $04.70.\_s$,\: $98.62.Dm$,\: $04.60.\_m,\: 14.70.Bh,\: 04.80.Nn$}
\keywords{Time, Modified gravity theory, Two-body problem, Rotation curves, Black holes, Photon, Quantum gravity, Gravitational redshift, Gravitational waves}
\maketitle
\section{Introduction}
Periodic relativity (PR) is a theory of accelerations in the flat space time. This article is a revised, improved and extended version of the previous work \cite{44,60}. In PR, the deviation to the flat Minkowski metric in the presence of the gravitational field gets introduced in the form,
\begin{align}\label{1.17}
\left(\frac{d\tau}{dt}\right)^2=(1-n\beta^2),
\end{align}
where the deviation factor $n$ is directly introduced in the Lorentz transformation equation. Here $t$ is the coordinate time, $\tau$ the proper time of the orbiting body, $n$ is a unitless real number and $\beta=v/c$. The corresponding line element in polar coordinates is,
\begin{align}\label{1.18}
ds^2=c^2dt^2-ndr^2-nr^2d\theta^2-n(r^2\sin^2{\theta})d\phi^2.
\end{align}
In this theory the weak field approximation of general relativity (GR) is eliminated. The unitless real number $n$ can be defined as a ratio of the theoretical acceleration to the experimentally observed acceleration in a special case of an orbital motion. One of the simplest form of actually observed acceleration is $v^2/r$. Complex form of Lorentz invariant acceleration is discussed later. Here any constant value of $n$ gives a flat metric. This gives,
\begin{align}\label{1.1b}
d\tau=dt\left(1-\frac{nv^2}{2c^2}\right),
\end{align}
to the first order accuracy for small values of $v$ and $n$. As discussed later, the line element of Eq.~\eqref{1.18} satisfies Einstein's field equation for the empty space $R_{\mu \nu}=0$ for any constant value of $n$. In PR it is proposed that the proper time $d\tau$ of a massive object such as a planet has a definite connection with the Doppler shift of the massive particles of which the massive object is composed. This causes every two body system to deviate differently from the flat Minkowski metric. Weak field approximation of general relativity (GR) \cite{1,2,3} does not allow such freedom. \\   
\hspace*{5 mm} In PR the factor $\left(\cos{\psi}+\sin{\psi}\right)$ associated with elliptical trajectory and ignored by both Newton and Einstein introduces geodesic like trajectories. Angle $\psi$ is shown in Figs.~\ref{Fig.1} and~\ref{Fig.3}. The field equations in the presence of matter are derived using the energy conservation law and proper use of the relativistic mass. The weak equivalence principle (WEP) is replaced by the dynamic WEP which states that the gravitational mass is equal to the relativistic mass. The main effect of having different deviation factors for different two body systems is that the proper time interval predictions of GR and PR are different, specially for distant objects. This is where the two theories could be tested for the soundness of their underlying logic of the meaning of time. Unfortunately, this test capability does not exist at present time.

Similarly, De Broglie pilot wave theory \cite{5,6} can also explain Hafele-Keating experiment \cite{123}. When Cesium atom in an atomic clock (in aircraft) moves relative to the observer on the surface of the earth, it develops a resultant pilot wave with respect to the observer and in the direction of the velocity of the aircraft. The resultant pilot wave is the sum total of all the individual pilot waves of the constituent particles of the cesium atom. The resultant pilot wave suffers kinematic and gravitational frequency shift as per the known science. In atomic clocks the time is measured by measuring the frequency of the hyperfine splitting of the 6s electron in the outer shell of the Cesium atom. This is measuring of frequency of light spectra in simple language. The world time standard relates 1 second with 9192631770 cycles/sec. or Hertz frequency of the light spectra in atomic clock. The frequency shift of the resultant pilot wave of the Cesium atom also includes contributions from the change in frequency of the hyperfine splitting of the 6s electron in the outer shell responsible for causing time dilation. Thus time is periodic in nature.
   
\subsection{Relativistic invariant}

\hspace*{5 mm} The relativistic invariant \(s^2\) presented by Minkowski, Lorentz and Einstein relates two points in space-time by the expression,
\begin{align}\label{1.1}
s^2=x^2+y^2+z^2-c^2t^2.
\end{align}
Here \(x^2+y^2+z^2\) represents three dimensional space, $c$ is the velocity of light and $t$ the ordinary linear time as we generally know.
Einstein's relativity theory is founded upon this simple equation and a hypothesis based on the E$\ddot{o}$tv$\ddot{o}$s experiment which showed that the gravitational mass of a body is equal to its inertial mass. It is also well known that if we replace $c$ with velocity $v<c$ for other massive particles, Eq.~\eqref{1.1} no longer remain meaningful. One of the fundamental proposal of the present theory is to recognize continuity between the electromagnetic wave spectrum and the massive particle wave spectrum. Such argument can be supported if we can supplant Eq.~\eqref{1.1} by an equation which is not only applicable to velocity of light but also to velocity of all the other particles which travel at speeds less than that of light.

\subsection{Periodic invariant}
It is possible to propose one such equation which I call the periodic invariant. It can be written as
\begin{align}\label{1.2}
s^2=\lambda^2-V^2T^2,
\end{align}
where $\lambda = h/p$ \ is the associated de Broglie
 wave\-length\cite{5,6}, $V = c^2/v$ the phase velocity, $v$ the particle velocity, and $T$ the period of the wave. 
One can see that Eq.~\eqref{1.2} does satisfy light particles as well as other massive particles. If we replace ordinary particle velocity with that of light, we get $v=c$, $V=c$ and
\begin{align}\label{1.2a}
s^2=\lambda^2-c^2T^2, 
\end{align}
If we multiply Eq.~\eqref{1.2a} for light with a real number $n^2$ and set $(n \lambda)^2$ equal to the cartesian distance \(x^2+y^2+z^2\) and $(nT)^2$ equal to the linear time $t^2$, Eq.~\eqref{1.2a} becomes equivalent to Eq.~\eqref{1.1}. Therefore Eq.~\eqref{1.1} is 
a special case of Eq.~\eqref{1.2}. Eq.~\eqref{1.2a} implies that space-time is wavy and that time does not flow in one direction but is strictly a periodic or cyclic phenomenon. Eqs.~\eqref{1.1} and \eqref{1.2a} both behave identically in the absence of gravitational field, but in the presence of $g$ field, for astronomical distances, Eq.~\eqref{1.2a} remains null and Eq.~\eqref{1.1} yields time like geodesics. In PR, both light as well as massive particles always travel along null paths. This makes it difficult to solve mundane problems of macroscopic proportions. This is why it becomes necessary to introduce approximations in the form of linear time and linear euclidean distance in Eq.~\eqref{1.2} which permits time like and space like geodesics for addressing the problems involving complex structures. However for certain fundamental measurements such as gravitational redshift and deflection of light, Eq.~\eqref{1.2a} should yield more accurate results than that given by Eq.~\eqref{1.1} because the reality does not get compromised.\\
\hspace*{5 mm} We can also say that Eqs.~\eqref{1.1} and \eqref{1.2a} both behave identically even in the presence of gravitational field when the space-time interval involves atomic and sub-atomic distances. Thus the validity of the algebraic structure (Clifford and Lie Algebra) associated with Eq.~\eqref{1.1} and the related gauge and spinor groups of particle physics is maintained. It is only at astronomical distances that the difference between two equations become perceptible and can affect any local symmetry formalism based on diffeomorphism. Therefore the validity of Dirac equation \cite{4} is maintained with respect to Eqs.~\eqref{1.2} and \eqref{1.2a}. Same is true for the algebra of Lorentz transformation when the space-time interval involves atomic and sub-atomic distances.

We can write Eq.~\eqref{1.2} as
\begin{align}\label{1.3}
s^2 = (ch/cp)^2-(c^2/v\nu)^2,						
\end{align}
where $h$ is Planck's constant, $p$ the particle momentum and $\nu = 1/T$ is the frequency of the associated de Broglie wave. This period-frequency relation is the only fundamental and basic equation that relates the concept of time to the physical world in an objectively real manner. The relativistic invariant relates the space and time continuum on a macro cosmic scale. The periodic invariant does the same on a microcosmic scale. If we introduce the energy-momentum invariant 
\begin{align}\label{1.4}
E^2 = E^2_0 + (cp)^2.
\end{align}
in Eq.~\eqref{1.3}, we get,
\begin{align}\label{1.5}
s^2 = ((hc)^2/(E^2-E^2_0)) - (c^2/v\nu)^2,	
\end{align}			
where $E$ = total energy of the particle and $E_0 = m_0c^2$ is the rest energy of particle. Relativistic mass is little used by modern physicists. Notwithstanding the modern usage we will use $m$ for relativistic mass and $m_0$ for rest mass throughout the article.

\section{Quantum Invariant}  
\hspace*{5 mm}The invariant Eq.~\eqref{1.5} has a general form applicable to all de Broglie particles. In relativity, the vanishing of the invariant $s^2$ given by Eq.~\eqref{1.1} does not mean that the distance between two space-time points gets 
obliterated. It simply means that the two space-time points can be connected by a light signal in vacuum. The new invariant Eq.~\eqref{1.5}, however, can vanish in two different ways. First, in the characteristic relativistic sense implying that two points in space-time can be connected by a energy signal (which can be a light signal or a massive particle signal), and secondly in an absolute sense where both terms on the right also vanish individually like the Euclidean invariant. In the first case, we get the relation, 
\begin{align}\label{2.1}
(E^2-E^2_0)/\nu^2 = (h^2v^2)/c^2.				
\end{align}
Substituting the photon parameters $E_0 = 0$ and $v = c$ into Eq.~\eqref{2.1} gives the quantum hypothesis of Max Planck, $~E = h\nu$. This provides sufficient reason to declare that Eq.~\eqref{2.1} is a general form of Max Planck's quantum hypothesis 
applicable to both massless as well massive particles.\\ 
\hspace*{5 mm} Essentially there is no difference between the relativistic invariant Eq.~\eqref{1.1} and the invariant Eq.~\eqref{1.5}, other than the fact that the former defines the space-time continuum and the latter defines the energy-vibration continuum. The equivalence of both these continuums will become clear when we define the quantum invariant with the assumptions that, given sufficient energy, all particles having rest masses can disintegrate into particles with zero rest masses; and that all particles having zero rest masses will have a constant velocity in space regardless of the inertial frames of reference and equal to the velocity of light. These two assumptions would allow us to adopt the hypothesis that the creation begins
with a vibration of the primal energy. We can introduce the photon parameters $E_0 = 0$ and $v = c$ in Eq.~\eqref{1.5} to simulate the initial state of the universe. This gives,
\begin{align}\label{2.2} 	    		
s^2 = (hc/E)^2 - (c/\nu)^2.				
\end{align}
And since the path of a massless particle is a null geodesic, for $s^2 = 0$,   Eq.~\eqref{2.2} can be further simplified to a form which is independent of the law of propagation of light. We shall call this form the Quantum Invariant.
\begin{align}\label{2.3}  
s^2 = (h/E)^2 - (1/\nu)^2.					
\end{align}
The quantum invariant can vanish in an absolute sense when $E > E_p$ and 
$\nu > \nu_p$, where $E_p$ and $\nu_p$ are Planck energy and Planck frequency respectively. In this case when the particle tries to acquire more energy than the Planck energy, it will violate the Compton limit on the wavelength and thus the particle wave will collapse and will become perfectly motionless leaving no mass gap. Thus the space-time continuum connecting two points gets completely obliterated and the resulting sub-quantic medium resembles a singularity characterised by a perfectly motionless indivisible field which is infinite like empty space. This is the fundamental substance of the universe from which a specific finite excitation had arisen as a wave particle duality.
Such a singularity suggests a motionless field devoid of ripples capable of giving birth to energy which is always in motion. 

The singular motionless field can not be described as energy because there are no oscillations in it. Since the singular motionless field is not the energy, it does not gravitate. The aether of the earlier theories is "energy in waves" which can interact with the motion of the planet but the motionless singular field is undetectable because it does not interact with any form of energy. Therefore in PR the accelerated expansion of the universe takes place within the infinitude of the singular motionless field. With the vibration in the small fraction of the singular motionless field comes the periodic phenomenon. Therefore time begins with the first vibration. Concept of proper time assumes linear time and distance scales whereas the true nature of reality is founded upon non-linear periods and wavelengths of the subatomic particles. Nevertheless to deal with a compound wave of a massive object such as a planet is not as simple as analyzing an individual particle. Thus the concept of proper time is useful in such cases as an approximation.

\subsection{Real and Unreal in Physics}
PR deals with very fine distinctions between what is relative and what is absolute; between what is real (tangible existence) and what is unreal (imaginary having no existence). Absolute is that which does not submit to the laws of relativity. Absolute is defined as one without a second. So there is no multiplicity in it. Consequently it does not have a boundary, otherwise second something can exist beyond that boundary. Hence Absolute is infinite. If Absolute is real than it must have a tangible existence. The relativistic invariant of special relativity is based on linear space (Euclidean distance) and linear time, both are imaginary and therefore unreal. In PR we have defined periodic invariant which is based on wavelengths and periods of particle waves. Here wavelengths and periods are imaginary but energy and frequency are real. Based on energy and frequency, we have defined quantum invariant which does not have imaginary space-time. Entire universe is made up of these fundamental particles, so the only thing that exists in the universe is the vibrating energy. When the vibrations subside, you get the underlying fundamental substance of the universe which is perfectly motionless and indivisible one without a second. Energy and space-time cannot exist without motion. Indivisibility of the fundamental substance causes particle fields, wave-particle duality, entanglement and gravity. So fundamentally time is periodic in nature. In PR, we have also defined a line element based on linear time and linear space which is different than general relativity in that it eliminates the weak field approximation and introduces a deviation factor $n$ which can be a constant or a ratio of accelerations. This eliminates the need for potential energy term in quantum mechanics and makes it easy to introduce gravity in quantum mechanics. In Section~\ref{GW} on gravitational waves we have employed this method for deriving the gravitational-wave strain amplitude and explaining the Gamma-ray burst observed in association with GW150914. Black hole analysis in PR is very much simplified compared to Kerr Black Hole because PR is energy based theory and not a geometrical theory.  

\section{Quantum Energy Equation}
General form of Max Planck's quantum hypothesis \eqref{2.1} can be written in various alternate forms of which Eq.~\eqref{3.3} is the most familiar.
\begin{align}
E=\{E^2_0+h^2\nu^2\beta^2\}^{1/2},\label{3.1}\\				
E=m_0c^2\{1+\gamma^2\beta^2\}^{1/2},\label{3.2}\\			
E=m_0c^2(1-\beta^2)^{-1/2},\label{3.3}\\					
E=\pm\{(m_0c^2)^2+(mc^2)(mv^2)\}^{1/2},\label{3.4}			
\end{align}
where $m$ is the relativistic mass, $\beta=v/c$ and $\gamma=m/m_0=(1-\beta^2)^{-1/2}$.		
In PR, $\beta$ and $\gamma$ need not be constants, however, their instantaneous values are related with each other and with other parameters in the same manner as in special relativity.

\subsection{True force}
\hspace*{5 mm} In order to come up with a truly invariant relationship between force and energy, we shall differentiate Eq.~\eqref{3.1} w.r.t. time.
\begin{align}
\frac{dE}{dt}=\frac{d}{dt}\{E^2_0+h^2\nu^2\beta^2\}^{1/2}=\mathbf{vF},\label{3.10}\\	\mathbf{vF}=\frac{1}{2E}\left(2h^2\nu^2\beta\frac{d\beta}{dt}+2h^2\beta^2\nu\frac{d\nu}{dt}\right),\label{3.11}\\
\mathbf{vF}=\frac{1}{2E}\left(2E^2\frac{v}{c^2}\frac{dv}{dt}+2Eh\frac{v^2}{c^2}\frac{d\nu}{dt}\right),\label{3.12}
\end{align}
where $\mathbf{F}$ is a new form of Lorentz force which we shall call the true force and $\mathbf{v}$ the velocity vector. Eq.~\eqref{3.10} reduces to
\begin{align}\label{3.13}
\mathbf{v}\mathbf{F}=v\left(ma+\frac{hv}{c^2}\frac{d\nu}{dt}\right),	
\end{align}
where $a=dv/dt$
is the acceleration of the particle. From Eq.~\eqref{3.13} we can deduce that the 
change in the energy of the particle is associated with two different changes in
the state of the particle.
\begin{itemize}
	\item The change in the velocity of the particle.
	\item The change in the frequency of the associated de Broglie wave.
\end{itemize}
When the second aspect is neglected, the invariant relationship between the force and the energy is lost. With respect to the massive particles, this second term on the right is comparable to the Doppler effect. Hence we will call it the de Broglie effect; and since this second term also has the units of force, we shall call this new force the de Broglie force. This shows that the true force consists of a sum of two forces, the classical force and the de Broglie force.

Even though there are equations in Einstein's relativity for relating force and energy, it remains a fact that the relativity theory has failed to provide satisfactory quantification of force and energy. The principle reason in my opinion is the concept of time as adopted by the relativity theory which assumes that the time is linear and flow in one direction from past to present and from present to future. This prevailing concept of time moving in one direction is a self-imposed illusion of the mind, just like imagining a blue sky which in reality is colorless, or riding a marry-go-round while all the time thinking that we are moving forward. Other authors have arrived at similar conclusion by analyzing the block universe concept \cite{79}. Relativity no doubt unifies the space and the time continuum, but because of the adoption of the linear time scales, it becomes very convenient to compromise the invariance of force and energy.

Both the classical as well as the relativistic mechanics are founded upon the assumption that $d\nu/dt$ is always zero for calculations involving force and energy. This is the very reason for which general relativity has failed to provide satisfactory quantification of force and energy. So whether one should hold on to the concept of proper time and linearity of time scales or adopt the view that the reality is based on non-linear periods? The answer to this question may not be very simple, but one thing is certain that if we assume $d\nu/dt=0$, then the derivation of gravitational redshift discussed in Section~\ref{GR} would not be possible.

\section{Two-body problem and the gravitational field}
\subsection{Dynamic weak equivalence principle}
\hspace*{5 mm} In order to deal with the static spherically symmetric gravitational field produced by a spherically symmetric body at rest, we work in a single plane and base our formalism on following postulate which basically means that the orbital energy is constant and consists of sum of kinetic energy and gravitational potential energy. 
\begin{quote}
In empty space, the rate of change of kinetic energy of a particle is equal to the rate of change of potential energy influencing the particle.
\end{quote}
In classical two-body problem, the gravitational field gets introduced as a single central potential acting radially, while the transverse component is assumed absent. In PR we will introduce gravitational field in terms of two components of a force acting normal and tangential to the particle trajectory, rather than as a resultant central force acting radially. This will allow us to account for the curvature of the trajectory. Moreover PR is not opposed to many of the conclusions of special and general relativity such as mass energy equivalence, distinction between rest mass and the relativistic mass, perihelic precession formalism, quadrupole formalism etc. Hence we will introduce another modification to Newton's inverse square law on following grounds. It is well recognized that the light is affected by the central gravitational potential due to rest mass of a body just like any other massive orbiting body. Making use of this observation we conclude that the mass of the orbiting body in the inverse square law should be relativistic mass and not the rest mass. This is because photon does not have any rest mass but only kinetic energy which can be correctly represented in terms of relativistic mass. And this relativistic mass is a variable parameter in gravitational field and not a constant. We shall adhere to this principle even while discussing massive bodies in orbit which also have some kinetic energy. Proof of the correctness of this assumption is evident in the derivation of the orbital period derivative of a binary star discussed in \cite{20}. These two changes introduces two more variables in addition to the radial distance in the formalism of central potential. Hence it is not very straight forward to introduce the classical theories of gravitational potentials in PR. However, in PR also the central potential acts radially and the transverse component is assumed absent. Therefore when the second body has only radial motion as in the case of gravitational redshift of light, the two additional variables disappear and the potential reduces to classical Newtonian potential. However, this does not happen in case of bending of light.\\
\hspace*{5 mm} How does the assumption that the gravitational attraction exists between the relativistic masses reflect on the equivalence principle? It appears that this assumption does not violate any of the three equivalence principles, the weak (WEP), Einstein (EEP) and the strong (SEP).
"Universality of free fall" is based on Newton's weak equivalence principle (WEP) which states: "the property of a body called mass is proportional to the weight." In WEP Newton did not specify whether the mass is the rest mass or relativistic mass
and no discussion of motion. In Einstein's notion, free fall indicates
inertial mass as well as relativistic mass but again in consideration of
E$\ddot{o}$tv$\ddot{o}$s experiment only inertial mass is equated to gravitational mass.
Besides, E$\ddot{o}$tv$\ddot{o}$s experiment is a static experiment which does not involve moving masses.
The present theory conforms with E$\ddot{o}$tv$\ddot{o}$s experiment when two gravitating bodies are in
equilibrium and at rest in the same coordinate system. For example, if an
object is thrown radially upwards in the coordinate system of the earth,
it will attain a maximum height and then freely fall back to the earth.
Momentarily when the object is at the maximum height, both the object and
the earth are perfectly at rest in the same coordinate system. At this
moment, the relativistic mass of the object is the same as its rest mass.
Therefore at this moment of equilibrium, the gravitational mass is equal
to inertial mass. Rest of the time it is the relativistic mass which is
equal to the gravitational mass. This is the dynamic version of WEP we have introduced in this theory which states that the gravitational mass of a body is equal to its relativistic mass. \\
\hspace*{5 mm} Whether one is working with the coordinate time of the central potential or with the proper time of the orbiting body, one of the two masses would certainly be the relativistic mass. So this factor needs to be considered. 
The effects of dynamic WEP gets absorbed in what is later defined as the deviation factor "$n$" and eventually shows up as a deviation in the proper time interval of the body. And proper time interval of the planet is not a part of the present day ephemerids \cite{66,67,68,69}. As long as the proper time interval is not included as one of the observables in ephemerids, there is no way to compare the GR predictions 
with the PR theory. The present day ephemerids are a three dimensional ephemerids. Introduction of proper time interval as a variable orbital parameter would introduce the fourth dimension to the ephemerids.

\begin{figure*}[ht]
\includegraphics[width=12cm]{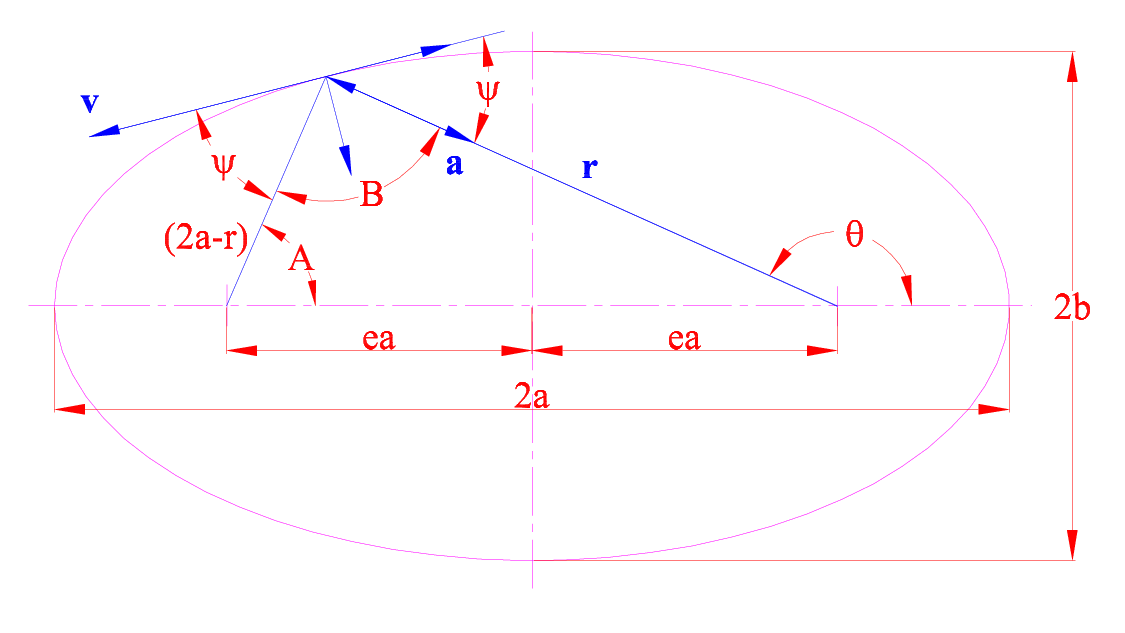}
\caption[]{(Color online) Two-body system\label{Fig.1}}
\end{figure*}

Fig.~\ref{Fig.1} shows the radial vector $\mathbf{r}$ connecting the central mass $M_0$ with the particle in motion having rest mass $m_0$. $\theta$ is the polar angle and $\psi$ is the angle between the radial vector $\mathbf{r}$ and the tangent vector $\wh{\mathbf{T}}$. Here we are dealing with two coordinate systems, one centered on the central mass with polar parameters and another centered on the particle in motion with axes along the tangent and the normal to the trajectory. Both these coordinate systems are oriented w.r.t. each other in such a way that the normal vector and the radial vector make an angle equal to $((\pi/2)-\psi)$ between them, and $\psi$ is a 
variable. Therefore in PR the transverse component is absent w.r.t. polar coordinate system of the central mass but not w.r.t. the coordinate system of the particle in motion. Hence, the rate of change of potential energy influencing the particle can be given by
\begin{align}\label{4.1b}
\frac{dE_p}{dt}=\mathbf{F}\cdot\mathbf{v}=m\mathbf{a}\cdot\mathbf{v}=,
\end{align}
\begin{align}\label{4.1}
-\frac{\mu m}{r^2}\left(\cos{\psi}+\sin{\psi}\right)\mathbf{\hat r}\cdot\mathbf{v}=-\frac{\mu m}{r^2}\mathbf{\hat r}\frac{d\mathbf{r}}{dt}\left(\cos{\psi}+\sin{\psi}\right).
\end{align}
where $\mu=GM_0$, $G=$gravitational constant, m = relativistic mass and following relations hold as usual. 
\begin{align*}
\mathbf{\hat r}\cdot\frac{d\mathbf{r}}{dt}=\mathbf{\hat r}\cdot\mathbf{v}=
\left(\frac{dr}{dt}\mathbf{\hat r}\cdot\mathbf{\hat r}+\frac{rd\theta}{dt}\boldsymbol{\hat \theta}\cdot\mathbf{\hat r}\right)=\frac{dr}{dt}.
\end{align*}
Gravitational potential can be deduced from Eq.~\eqref{4.1} as
\begin{align}\label{4.1a}
-\int \frac{\mu \gamma}{r^2} \left(\cos{\psi}+\sin{\psi}\right)dr,
\end{align}
where $\gamma=m/m_0$ is a variable. $\gamma$ and $\psi$ both are functions of $r$. 

\subsection{Massless particles in gravitational field}
\hspace*{5 mm} For massless particles, the rate of change of kinetic energy can be given by Eq.~\eqref{3.13} as described below. Following is applicable to all massless particles.
\begin{itemize}
\item	 The particles will have velocity equal to $c$. The particles cannot be accelerated along the direction of motion in a conventional sense. The wavefront can however be accelerated normal to the direction of motion. The wave can be subjected to Doppler shift along the direction of motion.
\end{itemize}

\subsubsection{Electromagnetic wave can be accelerated at constant velocity of photon}
Physicists usually argue that photon travel at a constant speed of light and cannot be accelerated but this is not true. Einstein's theory show that light can be subjected to gravitational redshift. In this, light is acted upon by gravitational acceleration at a constant velocity. In Newton's standard definition of force $F = ma$, the component of force causing acceleration of light is ignored. And with that additional component of acceleration causing the de Broglie force given in Eq.~\eqref{3.13}, the gravitational redshift of light can be calculated without using general relativity. Physicists usually define speed of light as distance traveled divided by time taken to travel the distance. But this is not very accurate definition. The correct definition is the wavelength divided by period of the electromagnetic wave. This ratio of wavelength to period is constant even when light experiences gravitational frequency shift (redshift) because with the change in the frequency (or period), the wavelength also change simultaneously in such a way that the velocity remains constant. In Periodic relativity, time is periodic so light can accelerate at constant velocity but in general relativity time is linear so light cannot accelerate at constant velocity. In general relativity if you accelerate the light, the velocity of light must increase, it cannot remain constant. This is because time dilation and length contraction does not occur simultaneously in general relativity, like the period and wavelength described above. Therefore gravitational redshift of light in general relativity itself is an indicator of failure of the equivalence principle. In periodic Relativity Newtonian force F =ma is modified as F = ma + de Broglie force. With this additional de Broglie force, how can you say gravitational mass is equal to inertial (rest) mass? This de Broglie force term shown in Eq.~\eqref{1.26} enables the explicit derivation of gravitation redshift of light. In PR, gravitational mass is equal to relativistic mass which solves this problem but relativistic mass is abandoned by general relativity theorists because it is not convenient to introduce weak field approximation in the flat Minkowski metric. This is the conflict between linear time and periodic time. At the quantum level, things don't work too well with the linear time. Therefore in general relativity the other three forces, electromagnetic, strong and weak can not cause time dilation and length contractions. But in periodic relativity they do. In PR we can also derive deflection of light without using general relativity. In this, we make use of angle $\psi$ shown in Eq.~\eqref{1.26} associated with the geometry of ellipse, which was ignored by both Newton and Einstein. Both these derivations satisfy Einstein's field equations for $n=1$.

\subsection{Curvilinear Gravity}
Newton realized the equality of gravitational force and the centrifugal force acting on the orbiting body and then introduced Kepler's third law of orbital periods to arrive at the inverse square law of gravitation given by
\begin{align}\label{1.1m}
m_0\frac{d^2\mathbf{r}}{dt^2}=
-\frac{GM_0m_0}{r^2}\mathbf{\hat r}.
\end{align}
where $GM_0=\mu$. Here we introduce the dynamic weak equivalence principle which states that the gravitational mass is equal to the relativistic mass. Therefore Eq.~\eqref{1.1m} becomes
\begin{align}\label{1.2m}
m\frac{d^2\mathbf{r}}{dt^2}=
-\frac{\mu m}{r^2}\mathbf{\hat r}.
\end{align}
In classical mechanics, we have two different expressions for the acceleration acting on a body in motion. One is a general expression $d\mathbf{v}/dt$ in cartesian coordinates which include the curvature term, and another is for Newtonian gravity in polar coordinates $d^2\mathbf{r}/dt^2$ based on the angular momentum vector $\mathbf{h}$, which is supposed to be a constant in order to satisfy Kepler's third law of equal areas in equal times. In Newton's theory the normal component of $d\mathbf{v}/dt$ containing curvature term is equated with radial acceleration $d^2\mathbf{r}/dt^2$. In PR we will show that these two accelerations are not equal. The reason for this is that the Newtonian gravity ignores the variation of angle $\psi$ along the trajectory by assuming constant $\mathbf{h}$. At the same time we maintain that the velocity vectors in both coordinate systems are equal, $\mathbf{v}=d\mathbf{r}/dt$. This is because $d\mathbf{r}/dt$ is not a radial vector. As shown in Figs.1, 2 and 3,  $\psi$ is the angle between the radial vector and the tangential velocity vector. Explanation given below makes it clear that the theory is Lorentz invariant and factor $\left(\cos{\psi}+\sin{\psi}\right)$ in Eq.~\eqref{4.1a} introduces geodesic like trajectories. The details are as follows. 

\begin{figure*}[ht]
\includegraphics[width=16cm]{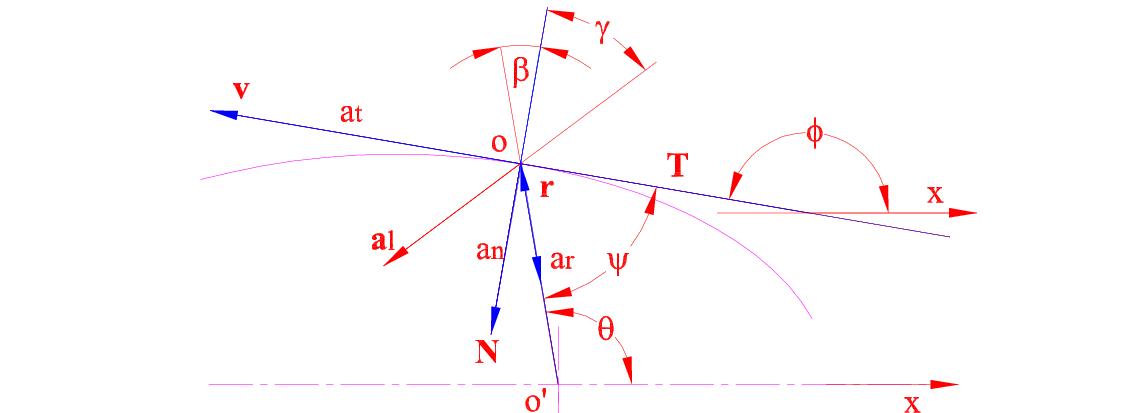}
\caption[]{(Color online) Vectors in a two-body system\label{Fig.2}}
\end{figure*}

\begin{figure*}[t]
\includegraphics[width=12cm]{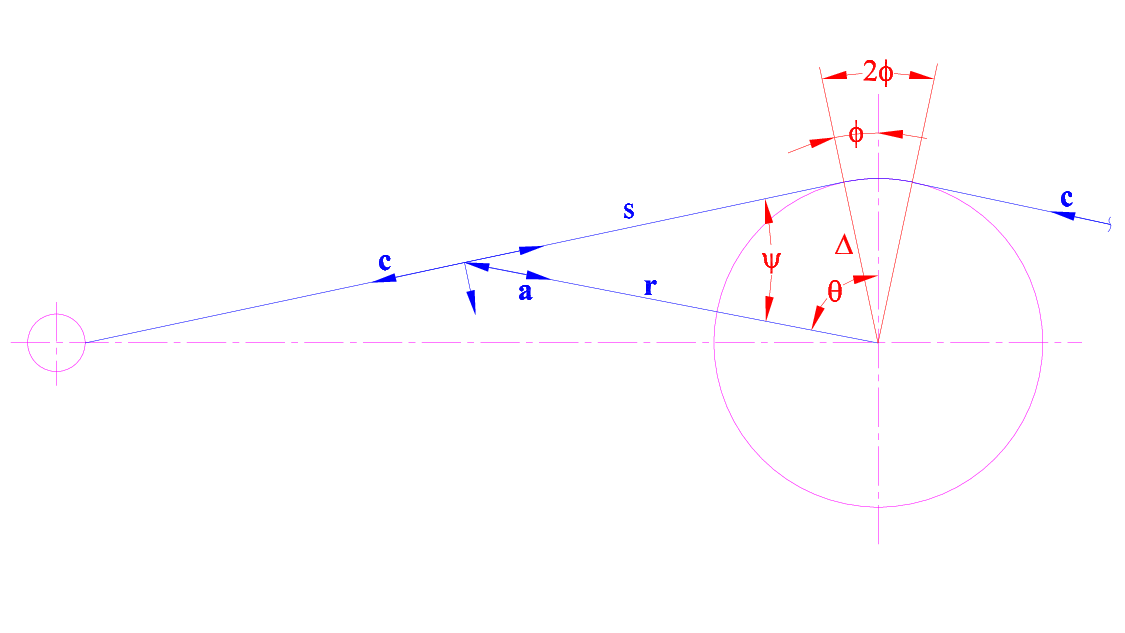}
\caption[]{(Color online) Bending of light around sun.\label{Fig.3}}
\end{figure*}

As shown in Fig.~\ref{Fig.2}, this angle $\psi$ is related to curvature through the expression 
\begin{align}\label{1.3m}
\phi=\theta+\psi.
\end{align}
where $d\phi/ds=\kappa$ is the curvature. Newtonian gravity ignores this curvature term by assuming constant $\psi=\pi/2$. This can be verified from following arguments.
\begin{align}\label{1.4m}
\mathbf{h}=\frac{\mathbf{L}}{m}=\frac{\mathbf{p}\boldsymbol{\times}\mathbf{r}}{m}\equiv\frac{|\mathbf{p}||\mathbf{r}|\sin{\psi}}{m}\mathbf{\hat h}=r^2\frac{d\theta}{dt}\sin{\psi}\mathbf{\hat h}.
\end{align}
From Eq.~\eqref{1.4m} we can see that $\mathbf{h}$ can be the desired constant only if $\sin{\psi}=1$. This shows that the Newtonian gravity approximates the curvature of the trajectory of the orbiting body. Hence in periodic relativity it is considered unreasonable to equate the normal component of the cartesian acceleration $d\mathbf{v}/dt$ with the Newtonian polar acceleration $d^2\mathbf{r}/dt^2$.

In order to account for the variation of angle $\psi$ along the trajectory, we propose that the absolute sum of vector and scalar products of $(\mu/r^2)\mathbf{\hat r}$ and $\mathbf{\hat a}$ is equal to magnitude of $d\mathbf{v}/dt$. The relation of these vectors to angle $\psi$ is shown in Fig.~\ref{Fig.2}.

\begin{align}\label{1.5m}
\left|\frac{d\mathbf{v}}{dt}\right|=\left|-\left|\mathbf{\hat a}\boldsymbol{\times}\frac{\mu}{r^2}\mathbf{\hat r}\right|-\frac{\mu}{r^2}\mathbf{\hat r}\cdot\mathbf{\hat a}\right|.
\end{align}
\begin{align}\label{1.6}
\left|\frac{d\mathbf{v}}{dt}\right|=\left||\mathbf{\hat a}|\left|\frac{\mu}{r^2}\mathbf{\hat r}\right|\sin{(\beta+\gamma)}\mathbf{\hat h}\right|+\left|\frac{\mu}{r^2}\mathbf{\hat r}\right||\mathbf{\hat a}|\cos{(\beta+\gamma)}.
\end{align}
where
\begin{align}\label{1.61}
\beta=\left(\frac{\pi}{2}-\psi\right).
\end{align}
\begin{align}\label{1.7}
\gamma=\tan^{-1}\left(\frac{a_t}{a_n}\right).
\end{align}
Various magnitudes of the parameters shown in Fig.~\ref{Fig.1} are as follows.
\begin{align}\label{1.8}
\mathbf{a}_l=\frac{d\mathbf{v}}{dt}.
\end{align}
\begin{align}\label{1.9}
a_t=\left(\frac{d^2s}{dt^2}+\frac{v}{\nu}\frac{d\nu}{dt}\right).
\end{align}
\begin{align}\label{1.10}
a_n=\kappa\left(\frac{ds}{dt}\right)^2.
\end{align}
\begin{align}\label{1.11}
a_r=-\frac{\mu}{r^2}=\left| \frac{d^2\mathbf{r}}{dt^2} \right|.
\end{align}
\begin{align}\label{1.12}
\mathbf{v}=\frac{d\mathbf{r}}{dt}.
\end{align}
Substitution of Eq.~\eqref{1.61} in Eq.~\eqref{1.6} gives
\begin{align}\label{1.13}
\left|\frac{d\mathbf{v}}{dt}\right|=\frac{\mu}{r^2}\left(\cos{(\psi-\gamma)}+\sin{(\psi-\gamma)}\right).
\end{align}
When the tangential component of the acceleration is absent then we have $a_t\mathbf{\wh T}=0$. This gives $\gamma=0$ and Eq.~\eqref{1.13} reduces to
\begin{align}\label{1.14}
\left|\frac{d\mathbf{v}}{dt}\right|=\frac{\mu}{r^2}\left(\cos{\psi}+\sin{\psi}\right).
\end{align}
Similarly we can show that
\begin{align}\label{1.15}
\left|\frac{d\mathbf{v}}{dt}\right|=\left| \frac{d^2\mathbf{r}}{dt^2} \right| \left(\cos{(\psi-\gamma)}+\sin{(\psi-\gamma)}\right).
\end{align}
The first term on the right of Eq.~\eqref{1.13} can be interpreted as an angular acceleration vector with its axis perpendicular to the plane of motion. This could be the additional acceleration quantity responsible for the rotation of the velocity vector $\mathbf{v}$ about the coordinate origin $o$, causing the curvature of the trajectory. 

\subsection{Lorentz invariant acceleration}
In relativity we can either write our equations in terms of proper time or alternatively we can write them in terms of relativistic mass. Eq.~\eqref{1.17} can be written as 
\begin{align}\label{1.19}
\left(\frac{d\tau}{dt}\right)^2=(1-n\beta^2)=\left(\frac{m_0}{m}\right)^2=\left(\frac{E_0}{E}\right)^2,
\end{align}
where $E=mc^2=h\nu$. This gives
\begin{align}\label{1.20}
E=(E_0^2+nE^2\beta^2)^{1/2}.
\end{align}
If we introduce photon parameters $m_0=0$ and $v=c$ in Eq.~\eqref{1.20}, we can recover photon energy $E=mc^2=h\nu$ by putting $n=1$ for flat Minkowski space time. Further we can write Eq.~\eqref{1.17} for photon as 
\begin{align}\label{1.17a}
\left(\frac{cd\tau}{cdt}\right)^2=(1-n),
\end{align}
\begin{align}\label{1.17b}
\left(\frac{ds}{cdt}\right)^2=(1-n),
\end{align}
For $n=1$ we get $ds^2=0$ signifying null trajectory for light.

Differentiating Eq.~\eqref{1.20} w.r.t. time for $n=1$ we get,
\begin{align}\label{1.21}
\left(\frac{1}{v}\right)\frac{dE}{dt}=\mathbf{\hat v}F=\left(m\mathbf{a}+\frac{h\mathbf{v}}{c^2}\frac{d\nu}{dt}\right).
\end{align}
Here we arrive at the same relation that we described as true force in Eq.~\eqref{3.13} except that now we have introduced the deviation factor $n$. I like to point out that this true force is same as the Lorentz force. Here we have used the relation $E=mc^2=h\nu$. Therefore 
\begin{align}\label{1.22}
\mathbf{F}=\frac{d\mathbf{p}}{dt}=\left(m\mathbf{a}+\frac{h\mathbf{v}}{c^2}\frac{d\nu}{dt}\right),	
\end{align}
where $\mathbf{F}$ is the Lorentz force and $\mathbf{v}$ the velocity vector and $\mathbf{a}$
is the classical acceleration of the particle given by
\begin{align}\label{1.23}
\mathbf{a}=\left(\frac{d^2s}{dt^2}\mathbf{\wh T}+\kappa\left(\frac{ds}{dt}\right)^2\mathbf{\wh N}\right).
\end{align}
Therefore,\\
Lorentz force = Classical force + de Broglie force.\\
From Eq.~\eqref{1.22} we can define Lorentz invariant acceleration $\mathbf{a}_l$ as
\begin{align}\label{1.24}
\mathbf{a}_l=\left(\left(\frac{d^2s}{dt^2}+\frac{v}{\nu}\frac{d\nu}{dt}\right)\mathbf{\wh T}+\kappa\left(\frac{ds}{dt}\right)^2\mathbf{\wh N}\right).
\end{align}
The de Broglie force acts along the tangent vector.
Now we equate Lorentz force with the gravitational force given by Eq.~\eqref{1.13}
\begin{align}\label{1.25}
\begin{split}
m\mathbf{a}_l&=\frac{d\mathbf{p}}{dt}=m\left(\left(\frac{d^2s}{dt^2}+\frac{v}{\nu}\frac{d\nu}{dt}\right)\mathbf{\wh T}+\kappa\left(\frac{ds}{dt}\right)^2\mathbf{\wh N}\right)\\
&=\left(0\mathbf{\wh T}+\frac{\mu m}{r^2}\left(\cos{(\psi-\gamma)}+\sin{(\psi-\gamma)}\right)\mathbf{\wh N}\right).
\end{split}
\end{align}
\begin{align}\label{1.26}
\begin{split}
\mathbf{a}_l&=\frac{d\mathbf{v}}{dt}=\left(\left(\frac{d^2s}{dt^2}+\frac{v}{\nu}\frac{d\nu}{dt}\right)\mathbf{\wh T}+\kappa\left(\frac{ds}{dt}\right)^2\mathbf{\wh N}\right)\\&=
\left(0\mathbf{\wh T}+\frac{\mu}{r^2}\left(\cos{(\psi-\gamma)}+\sin{(\psi-\gamma)}\right)\mathbf{\wh N}\right).
\end{split}
\end{align}
Eq.~\eqref{1.26} replaces the principle of equivalence of general relativity.

\subsubsection{Gravitational redshift in periodic relativity}\label{GR}
In PR the light is a wave and does not have any inertial mass, only the relativistic mass which is equivalent to its kinetic energy. We can apply Eq.~\eqref{1.26} to the gravitational redshift problem \cite{9,11,16,17} involving a photon traveling in a straight line from the sun to the earth along a path connecting their centers. In this case we have $d^2s/dt^2=0$, $\kappa=0$ and $\psi=0$, $\gamma=0$ and both, the gravitational force and the de Broglie force act along the radial direction, thus 
\begin{align}\label{4.3}
\frac{\mu m}{r^2}\mathbf{\hat r}=\frac{mv}{\nu}\frac{d\nu}{dt}\mathbf{\hat r}.
\end{align}
\begin{align*}
\frac{\mu m}{r^2}=\frac{h}{c}\frac{d\nu}{dt}=
\frac{h}{c}\frac{d\nu}{dr}\frac{dr}{dt}=h\frac{d\nu}{dr},
\end{align*}
\begin{align*}
\frac{\mu}{r^2}=\frac{hc^2}{mc^2}\frac{d\nu}{dr},\qquad\qquad  
\frac{\mu}{r^2}dr=\frac{h}{E}c^2d\nu = c^2\frac{1}{\nu}d\nu. 
\end{align*}
Integration over the entire trajectory gives
\begin{align}\label{4.4}
\int_\Delta^{\Delta+l}\frac{\mu}{r^2}dr=c^2\int_{\nu_s}^{\nu_\infty}\frac{1}{\nu}d\nu,
\end{align}
where $\Delta$ is the solar radius, $l$ the distance traveled by photon, $\nu_s$ the frequency of light on the surface of the sun and $\nu_\infty$ the frequency of light on earth. This gives
\begin{align}\label{4.5}
\frac{\mu}{c^2}\frac{l}{(\Delta^2+\Delta l)}=\frac{\varphi_1-\varphi_2}{c^2}=\ln\left(1-\frac{\delta\nu}{\nu_s}\right),
\end{align}
\begin{align}\label{grs1}
-\frac{\delta\nu}{\nu_s}=\frac{1}{c^2}(\varphi_1-\varphi_2)+\frac{1}{c^4}(\frac{1}{2}\varphi_1^2-\varphi_1\varphi_2+\frac{1}{2}\varphi_2^2)+\mathcal{O}\left(\frac{1}{c^6}\right).
\end{align}
The first order term is exactly the same value predicted by general relativity in terms of gravitational potentials $\varphi_1$ and $\varphi_2$ at locations separated by distance $l$, and verified experimentally \cite{9,16,17,11} with a high level of accuracy. The second order term for general relativity is slightly different,
\begin{align}\label{gr1}
+\frac{1}{c^4}(-\frac{1}{2}\varphi_1^2-\varphi_1\varphi_2+\frac{3}{2}\varphi_2^2),
\end{align}
and below the accuracy \cite{11,24,31,32,1} in the measurement of gravitational redshift.
For $l>100\Delta$, Eq.~\eqref{4.5} approaches
\begin{align}\label{4.6}
-\frac{\delta\nu}{\nu_s}=\frac{\mu}{c^2\Delta}.
\end{align}
For $l<0.01\Delta$, Eq.~\eqref{4.5} approaches the formula for the Doppler effect \cite{9}, given by
\begin{align}\label{4.7}
-\frac{\delta\nu}{\nu_s}=\frac{\mu l}{c^2\Delta^2}=\frac{gl}{c^2}=\frac{\vartheta}{c}.
\end{align}
It should be noted here that if Eqs.~\eqref{1.2} and \eqref{1.2a} had nothing to do with Eq.~\eqref{1.1}, then it would be impossible to derive the gravitational redshift Eq.~\eqref{4.5}. It is mainly due to the periodic representation of Eq.~\eqref{1.2} that the frequency term appears in Eqs.~\eqref{2.1}, ~\eqref{3.1}, ~\eqref{3.13}, and ~\eqref{4.3}. This makes it possible to derive Eq.~\eqref{4.5} without mentioning linear time or linear distance and without utilizing Riemannian geometry and geodesic trajectories. Hence unlike GR, the gravitational frequency shift in PR is invariant.\\
\hspace*{5 mm}The special theory of relativity assumes global coordinate systems and global invariance of speed of light as well as the equivalence of mass and energy. Einstein had to abandon the global coordinates and global invariance of speed of light while formulating general relativity because they were in conflict with the principle of equivalence of inertial mass and gravitational mass. However, he permitted the existence of a local system of inertial coordinates in a small region around any event. In PR we have gone one step further and restricted the local system of inertial coordinates to have only instantaneous existence. This makes the proper time a continuously variable phenomenon, which could now be identified with the continuously variable period of the body associated with the inertial coordinate. This has significant effect if the body is a fundamental particle such as a photon. When such instantaneous coordinate system is fixed on a massive body such as a planet, it acts exactly like the coordinate system of general relativity because the period of the associated wave of the planet does not change significantly from instance to instance and such is also the case with its proper time which remains practically constant. Similarly any fundamental particle traveling along a constant radial distance from a central massive body shall also have constant period and constant proper time. This is consistent with the general relativity definitions of tangential and radial velocities of light in a gravitational field \cite{1} given in geometrical units ($c_0 = 1$) by ,
\begin{align}\label{4.7a}
c_t=1+\varphi=\sqrt{1+2\varphi}=\sqrt{c_r}.
\end{align}
In general relativity, the rate of proper time at a fixed radial position in a gravitational field relative to the coordinate time can be obtained from general form of the metrical space time line element for a spherically symmetrical static field in polar coordinate and is given by
\begin{align}\label{4.7b}
\frac{d\tau(r)}{dt}=\sqrt{g_{tt}(r)}.
\end{align}
From Schwarzschild metric we have $g_{tt}(r)=1+2\varphi$. Now in general relativity, there is no explicit derivation of formula for gravitational redshift, but it is implicitly deduced from Eq.~\eqref{4.7b} and given by
\begin{align}\label{4.7c}
\frac{\nu_2}{\nu_1}=\frac{\sqrt{1+2\varphi_1}}{\sqrt{1+2\varphi_2}}.
\end{align}
The only support this formula has is the experimental verification of the first order term. Hence we would not be violating any scientific law if we propose that the correct implication of Eq.~\eqref{4.7b} is
\begin{align}\label{4.7d}
\frac{\nu_2}{\nu_1}=\frac{e^{\sqrt{1+2\varphi_1}}}{e^{\sqrt{1+2\varphi_2}}}.
\end{align}
Eq.~\eqref{4.7d} also yields the same first order term, besides it can also be explicitly derived and is exactly the same formula (in geometrical units) given by Eq.~\eqref{4.5}. It is to be noted here that Eq.~\eqref{1.2} and Eq.~\eqref{1.2a} has this unique property that they remain null even in a gravitational field. This is because equations contain only periodic time and wavelengths which are not independent of each other. This is not the case with Eq.~\eqref{1.1} which uses linear time and linear distance, so their interdependence is lost which is then restored with a deviation factor. In general relativity this deviation factor is used for all massive particles but is dismissed for photon on the ground that they travel along null geodesics. So computing proper time of photon is out of question in GR. But in PR we can use deviation factor $n$ for photon as well and compute proper time of photon which can be negligibly small value close to null. For this specific case we define the deviation factor $n$ as a ratio of Newtonian gravitational acceleration to the de Broglie acceleration given in Eq.~\eqref{4.3}. 
\begin{align}\label{4.3a}
n=\frac{GM}{r^2}\left(\frac{h}{mc}\frac{d\nu}{dt}\right)^{-1}
\end{align}
If we look at Eq.~\eqref{1.14}, we find that Eq.~\eqref{4.3a} is valid for both cases $\psi=0$ and $\psi=\pi/2$.
When this ratio is $n=1$, we get null geodesics but for any radial motion to take place, this ratio has to deviate from 1, otherwise the particle will make only circular motion due to equal and opposite radial forces acting on it. So for radial motion to take place, we must have $n=(1\pm \epsilon)$ where $\epsilon$ can be a very small value close to zero and $\pm$ sign depends on whether the trajectory is time like or space like. In Eq.~\eqref{1.17}, if we substitute $v=c$ and  $n=(1\pm \epsilon)$, we get
\begin{align}\label{4.7e}
\left(\frac{d\tau}{dt}\right)^2=\mp \epsilon.
\end{align}
The positive sign on the right should give time like geodesic applicable to electromagnetic waves and negative sign gives space like geodesic applicable to gravitational waves. Both travel at speed of light. 

\subsubsection{Bending of light in periodic relativity}
For the bending of light around the sun \cite{18,11,12}, we introduce light parameters $v=ds/dt=c$, $d^2s/dt^2 =0$ and $cdt=ds$, along with $\kappa=d\phi/ds$ for the curvature of the trajectory in Eq.~\eqref{1.26}. In this case we will have $d\nu/dt=0$ because the ray is equally blue shifted and then red shifted, and the frequency shift is $0$ at the limb of the sun.
This gives,
\begin{align}\label{4.10}
\left|\frac{c^2}{\nu}\frac{d\nu}{ds}\mathbf{\wh T}+c^2\frac{d\phi}{ds}\mathbf{\wh N}\right|=\left|\frac{\mu}{r^2}\left(\cos{(\psi-\gamma)}+\sin{(\psi-\gamma)}\right)\mathbf{\wh N}\right|.
\end{align}
Multiplying both sides by $d\psi$, we get
\begin{align}\label{4.11}
\begin{split}
&\left|\frac{1}{\nu} d\nu d\psi \mathbf{\wh T}+ d\phi d\psi \mathbf{\wh N}\right|
=\frac{\mu}{c^2r^2}\left(\cos{(\psi-\gamma)}+\sin{(\psi-\gamma)}\right)ds d\psi.
\end{split}
\end{align}
We integrate both sides with proper limits. For the star light approaching the sun we get,
\begin{align}\label{4.12}
\begin{split}
&\left|\int_{\nu_1}^{\nu_2} \int_\pi^{\frac{\pi}{2}}\frac{1}{\nu}d\nu d\psi \mathbf{\wh T}+\int_{-\phi}^0 \int_\pi^{\frac{\pi}{2}}d\phi d\psi \mathbf{\wh N}\right|
=\frac{\mu}{c^2}\int_{-\infty}^0 
\int_\pi^{\frac{\pi}{2}} \frac{1}{r^2}\left(\cos{(\psi-\gamma)}+\sin{(\psi-\gamma)}\right)d\psi ds.
\end{split}
\end{align}
For the star light approaching earth from the limb of the sun we get,
\begin{align}\label{4.13}
\begin{split}
&\left|\int_{\nu_2}^{\nu_1} \int_{\frac{\pi}{2}}^0\frac{1}{\nu}d\nu d\psi \mathbf{\wh T}+\int_0^{-\phi} \int_{\frac{\pi}{2}}^0 d\phi d\psi \mathbf{\wh N}\right|
=\frac{\mu}{c^2}\int_0^{\infty} 
\int_{\frac{\pi}{2}}^0 \frac{1}{r^2}\left(\cos{(\psi-\gamma)}+\sin{(\psi-\gamma)}\right)d\psi ds.
\end{split}
\end{align}
\begin{align}\label{4.14}
\begin{split}
&\left|(\text{ln}\nu_2-\text{ln}\nu_1)\mathbf{\wh T}+\phi \mathbf{\wh N}\right|
=\frac{\mu}{c^2}\int_{-\infty}^0 
\int_\pi^{\frac{\pi}{2}} \frac{1}{r^2}\left(\cos{(\psi-\gamma)}+\sin{(\psi-\gamma)}\right)d\psi ds.
\end{split}
\end{align}
\begin{align}\label{4.15}
\begin{split}
&\left|(\text{ln}\nu_1-\text{ln}\nu_2)\mathbf{\wh T}+\phi \mathbf{\wh N}\right|
=\frac{\mu}{c^2}\int_0^{\infty} 
\int_{\frac{\pi}{2}}^0 \frac{1}{r^2}\left(\cos{(\psi-\gamma)}+\sin{(\psi-\gamma)}\right)d\psi ds.
\end{split}
\end{align}
If we add l.h.s. of Eqs.~\eqref{4.14}\ and~\eqref{4.15} we get,
\begin{align}\label{4.16}
l.h.s. = \left|0.\mathbf{\wh T}+2\phi \mathbf{\wh N}\right|.
\end{align}
From Eq.~\eqref{4.16} we see that the magnitude of the tangential component is zero. Therefore $\gamma=0$. Hence substituting $r^2=s^2+\Delta^2$ in Eqs.~\eqref{4.14} and~\eqref{4.15} we get
\begin{align}\label{4.17}
2\phi=\frac{4\mu}{c^2\Delta}.
\end{align}
We have used Eqs.~\eqref{1.21} and~\eqref{1.23} to explain both, the bending of light and the gravitational frequency shift, which correspond to the flat Minkowski metric with $n=1$. As can be seen, the higher order terms does not exist in PR theory. The second order term in general relativity is below the accuracy in the measurement of deflection of light \cite{18,11,24,1}, and is given by
\begin{align}\label{gr2}
+\left(\frac{15\pi}{4}-4\right)\left(\frac{\mu}{c^2\Delta}\right)^2.
\end{align}
\hspace*{5 mm}Experimental verification of the second order effect given by Eq.~\eqref{gr2} was the principal goal of LATOR mission \cite{23} which is now abandoned. If this theory is correct, the experiment can yield null result. If the fundamental postulates of a theory, physical or mathematical, are built upon approximations, then there is a chance of appearance of pseudo terms resembling higher order terms in the end results. We have a reason to believe that the orbital energy equations~\eqref{1.21} and~\eqref{1.23} in PR are exact in nature, and that is not the case with general relativity. In general relativity, Newtonian potential gets introduced into metric component $g_{00}$ as a deviation to flat Minkowski metric. This constitutes the weak-field approximation. For this very reason Schwarzschild metric does not remain null for light in the gravitational field as we have already discussed. Other competing theories modify Newtonian potential by way of Poisson's equation and multipole expansion. Another significant approximation in Schwarzschild solution is the assumption that the angle of deflection is subtended at the center of the sun. In PR, even though the schematic diagram shows the same arrangement, the calculations give us actual angle $\phi$ measured between the line $\theta=0$ and the line normal to the velocity vector at the end of the trajectory. These factors add up to give different second order terms in these two theories. It is interesting to note that the higher order terms are in higher powers of Newtonian potential. It should also be noted that we have utilized Eq.~\eqref{1.2} for the slow moving accelerating particle for determining the bending of light.

\subsection{Curvic and conic gravity}
Newtonian gravity is based on the constant vector $\mathbf{h}$ which yields the conic sections. Therefore we can distinguish the gravity that uses the Lorentz invariant acceleration as the curvilinear (or curvic) gravity and the Newtonian gravity with constant $\mathbf{h}$ as the conic gravity. Accelerations of the curvic and conic gravity are related by Eq.~\eqref{1.15}.

It also needs to be understood that $d^2\mathbf{r}/dt^2$ is a radial vector but $d\mathbf{r}/dt$ is not a radial vector which acts along the velocity vector $\mathbf{v}$. Moreover, the constant vector $\mathbf{h}$ does not play any role in defining the velocity vector $\mathbf{v}$. Therefore factor $\left(\cos{\psi}+\sin{\psi}\right)$ does not appear in this expression of velocity $\mathbf{v}=d\mathbf{r}/dt$ which remains unaltered. This can be verified from following analysis. By definition we have
\begin{align}\label{1.27}
\cos{\psi}=\frac{dr}{ds},\ \ \ \text{and}\ \ \   \sin{\psi}=\frac{rd\theta}{ds}.
\end{align}
\begin{align}\label{1.28}
\frac{d\mathbf{r}}{dt}=
\left(\frac{dr}{dt}\mathbf{\hat r}+\frac{rd\theta}{dt}\boldsymbol{\hat \theta}\right).
\end{align}
\begin{align}\label{1.29}
\frac{d\mathbf{r}}{dt}=
\frac{ds}{dt}\left(\cos{(\psi+\theta)}\mathbf{i}+\sin{(\psi+\theta)}\mathbf{j}\right).
\end{align}
Substitution of Eq.~\eqref{1.3} gives
\begin{align}\label{1.30}
\frac{d\mathbf{r}}{dt}=
\frac{ds}{dt}\sqrt{\left(\cos^2{\phi}+\sin^2{\phi}\right)}\mathbf{\wh T}=\frac{ds}{dt}\mathbf{\wh T}=\mathbf{v}.
\end{align}
From Fig.~\ref{Fig.1} we can verify that the unit vector acting at an angle $\phi$ is $\mathbf{\wh T}$. Therefore 
Eq.~\eqref{1.30} is not influenced by the constant $\mathbf{h}$ assumption.

\subsection{Massive particles in gravitational field}
\hspace*{5 mm} From Eqs.~\eqref{1.26} and \eqref{3.13} we get,
\begin{align}\label{4.13m}
\begin{split}
a(-\mathbf{\hat r})=\left(\frac{d^2s}{dt^2}\mathbf{\wh T}+\kappa\left(\frac{ds}{dt}\right)^2\mathbf{\wh N}\right).
\end{split}
\end{align}
\begin{align}\label{4.14m}
\frac{d\mathbf{v}}{dt}=\left(\kappa\left(\frac{ds}{dt}\right)^2\mathbf{\wh N}\right)=\frac{d^2\mathbf{r}}{dt^2}\left(\cos{\psi}+\sin{\psi}\right)\mathbf{\wh N}.
\end{align}
So any conversion of acceleration between radial direction and normal to tangential velocity vector is accompanied by the conversion factor $\left(\cos{\psi}+\sin{\psi}\right)$. This factor acts as a single scalar quantity and does not get split into normal and tangential vector components.

\subsubsection{Perihelic precession of planets}\label{ppp}
We assume that the general relativity theory as applicable to solar system planets is valid in weak-field approximation. We also declare however, that the theory fails to predict accurate higher order terms for gravitational red-shift and deflection of light, because in introducing the weak-field approximation, it compromises the global invariance of speed of light in gravitational field. The weak-field approximation also leads to the value of limiting radius of the event horizon of a black hole which is at variance with the measured value for M87 black hole. While Schwarzschild solution is sufficiently accurate in predicting the perihelic precession \cite{1,10,13} of the planets of the solar system, it may not be dependable for describing the photon trajectories in strong gravitational fields. Not only so, even trajectories of massive bodies in a strong gravitational field or extremely weak-gravitational field compared to Sun can deviate significantly from the Schwarzschild solution. In PR we can present a very simple and accurate derivation for perihelic precession of planets as given below, compared to which the general relativity derivation \cite{19,14,15} is very complicated. In PR from  Eqs.~\eqref{1.13} and ~\eqref{1.15} we get,
\begin{align}\label{1.15a}
\begin{split}
\frac{d\mathbf{v}}{dt}&=\left|\frac{d^2\mathbf{r}}{dt^2}\right|\left(\cos{(\psi-\gamma)}+\sin{(\psi-\gamma)}\right)\mathbf{\hat N}\\ &=-\left|\frac{\mu}{r^2}\mathbf{\hat r}\right|\left(\cos{(\psi-\gamma)}+\sin{(\psi-\gamma)}\right)\mathbf{\hat N}.
\end{split}
\end{align}
Eq.~\eqref{1.15a} contains Newton's inverse square law of gravity given by 
\begin{align}\label{1.16a}
\frac{d^2\mathbf{r}}{dt^2}=-\frac{\mu}{r^2}\mathbf{\hat r}=
-\frac{GM_0}{r^2}\mathbf{\hat r}.
\end{align}
Now we introduce the line element of periodic relativity in Eq.~\eqref{1.16a}. The line element is obtained from,
\begin{align}\label{1.31}
\left(\frac{d\tau}{dt}\right)^2=(1-n\beta^2),
\end{align}
where $\beta=v/c$, and $n$ is a constant real number which keeps the Minkowski metric flat.
With this we replace the coordinate time interval $dt$ of Eq.~\eqref{1.16a} with the proper time interval $d\tau$. 
\begin{align}\label{1.32}
\frac{d^2\mathbf{r}}{d\tau^2}(1-n\beta^2)=-\left(\frac{\mu}{r^2}\right)\mathbf{\hat r}.
\end{align} 
\begin{align}\label{1.32a}
\frac{d^2\mathbf{r}}{d\tau^2}=-\left(\frac{\mu}{r^2}\right)(1-n\beta^2)^{-1}\mathbf{\hat r}.
\end{align} 
\begin{align}\label{1.32b}
\frac{d^2\mathbf{r}}{d\tau^2}=-\left(\frac{\mu}{r^2}\right)(1+n\beta^2)\mathbf{\hat r}.
\end{align} 
We will write $v^2$ on r.h.s. in polar coordinates. 
\begin{align}\label{1.17c}
\begin{split}
&\frac{d^2\mathbf{r}}{d\tau^2}=-\left(\frac{\mu}{r^2}\right)\mathbf{\hat r}-\left(\frac{n\mu}{r^2c^2}\right) \left[\left(\frac{dr}{dt}\right)^2+r^2\left(\frac{d\theta}{dt}\right)^2+r^2 \sin^2\theta \left(\frac{d\phi}{dt}\right)^2 \right]\mathbf{\hat r}.
\end{split}
\end{align}
Working in $(r,\theta)$ plane we can put $d\phi=0$ and on r.h.s. substitute 
\begin{align}\label{1.18a}
r^2\left(\frac{d\theta}{dt}\right)^2=\frac{h^2}{r^2}.
\end{align}
Hence Eq.~\eqref{1.17c} reduces to
\begin{align}\label{1.17d}
\frac{d^2\mathbf{r}}{d\tau^2}=-\left(\frac{\mu}{r^2}\right)\mathbf{\hat r}-\left(\frac{n\mu}{r^2c^2}\right)
\left[\left(\frac{dr}{dt}\right)^2+\frac{h^2}{r^2} \right]\mathbf{\hat r}.
\end{align}
On l.h.s. we substitute
\begin{align}\label{4.18}
\frac{d^2\mathbf{r}}{d\tau^2}=\left(\frac{d^2 r}{d\tau^2}-\frac{h^2}{r^3}\right)\mathbf{\hat r}.
\end{align}
\begin{align}\label{1.17e}
\left(\frac{d^2 r}{d\tau^2}-\frac{h^2}{r^3}\right)\mathbf{\hat r}=-\left[\left(\frac{\mu}{r^2}\right)+\left(\frac{n\mu}{r^2c^2}\right)
\left(\frac{dr}{dt}\right)^2+\left(\frac{n\mu h^2}{r^4c^2}\right)\right]\mathbf{\hat r}.
\end{align}
\begin{align}\label{1.17f}
\frac{d^2 r}{d\tau^2}=-\left(\frac{\mu}{r^2}\right)-\left(\frac{n\mu}{r^2c^2}\right)
\left(\frac{dr}{dt}\right)^2-\left(\frac{n\mu h^2}{r^4c^2}\right)+\frac{h^2}{r^3}.
\end{align}
On the l.h.s. of Eq.~\eqref{1.17f} we have the proper time and on the r.h.s. we have the coordinate time. To obtain the time independent solution of the equation of motion we make the following substitution which gives a second order non-homogeneous, non-linear differential equation.
\begin{align}\label{1.17g1}
u=\frac{1}{r}, \quad \text{and} \quad \frac{d}{d\tau}=\frac{d}{dt}=hu^2\frac{d}{d\theta}.
\end{align} 
\begin{align}\label{1.17g}
\begin{split}
-h^2u^2\frac{d^2 u}{d\theta^2}=&-\mu u^2-\left(\frac{n\mu h^2}{c^2}\right)u^4+h^2u^3 -\left(\frac{n\mu}{c^2}\right)u^2\left(-u\frac{du}{d\theta}\right)^2.
\end{split}
\end{align}
\begin{align}\label{1.17h}
\frac{d^2 u}{d\theta^2}=\frac{\mu}{h^2}+\left(\frac{n\mu }{c^2}\right)u^2-u+\left(\frac{n\mu}{c^2h^2}\right)\left(-u\frac{du}{d\theta}\right)^2.
\end{align}
Angle $\psi$ between the radial vector and the velocity vector  is defined as 
\begin{align}\label{4.40c}
\psi=\tan^{-1}{\frac{r}{\dot{r}}} \qquad where \qquad \dot{r}=\frac{dr}{d\theta}.
\end{align}
\begin{align}\label{4.40d}
\sin{\psi}=(r/\dot{r})/\sqrt{(r/\dot{r})^2+1}.
\end{align}
\begin{align}\label{4.40e}
\sin^2{\psi}=(r/\dot{r})/\sqrt{(r/\dot{r})^2+1}.
\end{align}
\begin{align}\label{4.40f}
\sin^2{\psi}=\frac{r^2}{r^2+\dot{r}^2}=\left[1+\frac{1}{u^2}\left(\frac{du}{d\theta}\right)^2\right]^{-1}.
\end{align}
Using Eq.~\eqref{4.40f} we can write,
\begin{align}\label{4.40g}
\left(\frac{du}{d\theta}\right)^2=\frac{u^2}{\sin^2{\psi}}-u^2.
\end{align} 
General relativity has ignored angle $\psi$ by assuming $\psi=\pi/2$. This gives,
\begin{align}\label{4.40h}
\left(\frac{du}{d\theta}\right)^2=0.
\end{align} 
Substituting Eq.~\eqref{4.40h} in Eq.~\eqref{1.17h} gives,
\begin{align}\label{1.17i}
\frac{d^2 u}{d\theta^2}+u=\frac{\mu}{h^2}+\frac{n\mu u^2 }{c^2}.
\end{align}
Eq.~\eqref{1.17i} is exactly the same equation obtained in general relativity for $n=3$. So rest of the derivation for perihelic precession of planets is same as in GR. In Section~\ref{Efe} we will show that line element of PR given by Eq.~\eqref{1.18} satisfies Einstein's field equations for all constant values of $n$.

\subsubsection{Proper time of a planet}
We substitute value of $n=3$ from Eq.~\eqref{1.17i} in Eq.~\eqref{1.1b} for proper time and for second order velocity term on the right we will introduce classical vis-viva equation for planetary velocity
\begin{align}\label{4.34a}
v^2=\mu\left(\frac{2}{r}-\frac{1}{a}\right).
\end{align}
\begin{align}\label{4.39a1}
d\tau=dt\left(1-\frac{n\mu}{2c^2}\left(\frac{2}{r}-\frac{1}{a}\right)\right).
\end{align}
Substituting $r=h^2/\mu(1+e\cos{\theta})$ to obtain 
\begin{align}\label{4.39a}
d\tau=dt\left(1-\frac{n\mu(1+2e\cos{\theta}+e^2)}{2c^2a(1-e^2)}\right).
\end{align}
For $n=3$ and for circular orbits $a=r$ and $e=0$. This gives
\begin{align}\label{4.39b}
d\tau=dt\left(1-\frac{3\mu}{2c^2r}\right).
\end{align}
GR does not have a convenient way of expressing proper time equation such as Eq.~\eqref{4.39a} but it does provide expression for proper time in equatorial Keplerian circular orbits which is exactly same as Eq.~\eqref{4.39b}. Unlike GR, in PR value of $n$ is locked by measurement of proper time.

\section{Black hole}\label{BH}
The line element of the periodic relativity Eq.~\eqref{1.17} introduces deviation to the flat Minkowski metric in the presence of the gravitational field without affecting flatness of the metric. We can introduce velocity $v$ in polar coordinates in Eq.~\eqref{1.17} as given below.
\begin{align}\label{2.2m}
\begin{split}
&\left(\frac{d\tau}{dt}\right)^2=\left[1-\left(\frac{n}{c^2}\right)\left\{\left(\frac{dr}{dt}\right)^2+r^2\left(\frac{d\theta}{dt}\right)^2+r^2 \sin^2\theta \left(\frac{d\phi}{dt}\right)^2 \right\}\right].
\end{split}
\end{align}
We can apply Eq.~\eqref{2.2m} to a black hole by working in $(r,\phi)$ plane and putting $\theta=\pi/2$ and $d\theta=0$.
\begin{align}\label{2.2n}
(c d\tau)^2=(c dt)^2-n(dr^2+r^2 d\phi^2).
\end{align} 
\begin{align}\label{2.2o}
\frac{1}{nr^2}\left(\frac{ds}{d\phi}\right)^2=\frac{c^2}{nr^2}\left(\frac{dt}{d\phi}\right)^2-\frac{1}{r^2}\left(\frac{dr}{d\phi}\right)^2-1.
\end{align} 
For light trajectory we have $ds^2=0$ and we substitute 
\begin{align}\label{2.3m}
r^2\left(\frac{d\phi}{dt}\right)^2=\omega^2r^2=\frac{h^2}{r^2},
\end{align}
where $h$ is a constant. This gives
\begin{align}\label{2.3n}
\frac{n}{c^2}\left[\frac{1}{r^2}+\left(\frac{1}{r^2}\frac{dr}{d\phi}\right)^2\right]=\frac{1}{\omega^2r^4}=\frac{E^2}{L^2c^4}=\frac{1}{h^2}=\frac{1}{\mu a(1-\epsilon^2)},
\end{align}  
where $a$ is the semi-major axis of ellipse and $\epsilon$ is the eccentricity and $\mu=GM$. The total black hole mass $M$, has contributions from the irreducible mass, $M_{irr}$ and the mass-energy associated with the spin angular momentum of the black hole $S=M^2c^2 \chi$, where rotation parameter $\chi \in [0,1]$ \cite{156,174,175,176}. The irreducible mass is given by
\begin{align}\label{2.3nn}
M_{irr}=M\sqrt{\frac{1+\sqrt{1-\chi^2}}{2}}.
\end{align} 
Thus for circular orbit with $dr/d\phi=0$, $\epsilon=0$ and $a=r$, $n$ in Eq.~\eqref{2.3n} has to satisfy the condition
\begin{align}\label{2.3o}
n=\frac{c^2/r}{\mu/r^2}=\frac{c^2 r}{\mu}=1.
\end{align} 
This assures that the gravitational acceleration acting on photon is balanced by the centrifugal acceleration acting on it due to the curvature of light trajectory $\kappa=1/r_\kappa=1/r$. This condition is satisfied only if we put $r=1R_g=\mu/c^2$. This defines the limiting radius of the event horizon of a black hole in PR. Earlier we have seen that for gravitational redshift, deflection of light around sun and for gravitational waves, $n=1$ has always worked. Secondly it also satisfies the equality of centrifugal and gravitational accelerations at the event horizon. In GR the ratio of accelerations at event horizon $(c^2r/\mu)$ is $1/2$ and $ds^2=\infty$ and not zero. So the GR equations fail here but the PR equations do not fail at the event horizon. In PR, $R_g$ is defined with total mass which is relativistic mass including spin energy. If irreducible mass is used for defining $R_g$ then the event horizon is between $1R_g$ and $\sqrt{2}R_g$. The event horizon of Kerr black hole is between $1R_g$ and $2R_g$.

\subsection{Shadow of a Black hole}
Eq.~\eqref{2.3n} reduces to
\begin{align}\label{2.3p}
n\left[\frac{1}{r^2}+\left(\frac{1}{r^2}\frac{dr}{d\phi}\right)^2\right]=\frac{E^2}{L^2c^2}=\frac{1}{b^2},
\end{align}  
where $b$ is a constant. From Eq.~\eqref{2.3p} we can see that a light ray with impact parameter $b$ can reach a point at a distance $r$ only if $b\leq r/\sqrt{n}$ \cite{172}. For $n=1$ we get $b\leq r$. So practically there is no limit on $b$ in PR like in GR. To explain the shadow of a black hole, we will have to introduce new concepts in PR which are not available to GR. In principle, we have understood from Eq.~\eqref{2.3o} that the deviation factor $n$ defines ratio of accelerations in PR. GR has ignored de Broglie acceleration associated with de Broglie force. De Broglie acceleration can accelerate light at a constant velocity $c$. So we will expand the expression for $n$ given by Eq.~\eqref{2.3o} to include the Lorentz invariant acceleration which we derived earlier in Eq.~\eqref{1.26} and equated it with the gravitational acceleration. In this equation the de Broglie force acts along the tangent vector. Deviation factor $n$ can then be defined as $n=l.h.s./r.h.s.=1$. So here  $n$ is not ad hoc but derived using PR formalism. For a ray of light, this equation reduces to 
\begin{align}\label{1.26a}
n=\frac{\left((c/\nu)(d\nu/dt)\mathbf{\wh T}+(c^2/r_\kappa)\mathbf{\wh N}\right)}
{\left(0\mathbf{\wh T}+(\mu/r^2)\left(\cos{(\psi-\gamma)}+\sin{(\psi-\gamma)}\right)\mathbf{\wh N}\right)}=1.
\end{align}
Here $r_\kappa$ is the radius of curvature of the light ray. We can select the initial condition such that $\gamma=0$ and if the energy of the light ray is constant then $d\nu/dt=0$. This gives geodesic like trajectory with
\begin{align}\label{2.11c}
n=\left(\frac{r^2}{\mu(cos\psi+\sin\psi)} \times \frac{c^2}{r_k}\right)=1,
\end{align}
where
\begin{align}\label{2.11cc}
(cos\psi+\sin\psi)=\left[\sqrt{1-\frac{h^2}{v^2r^2}}+\frac{h}{vr}\right].
\end{align}
For $\psi=\pi/2$ we get circular photon trajectory.
\begin{align}\label{2.11d}
n=\left(\frac{r^2}{\mu} \times \frac{c^2}{r_k}\right)=1.
\end{align}
This is the same Eq.~\eqref{2.3o} that we derived from the line element of PR Eq.~\eqref{2.2m}. If we define $r=xR_g$ where $x$ is a real number $> 0$ and the gravitational radius $R_g=\mu/c^2$, then Eq.~\eqref{2.11d} can be satisfied only if $r_k=x^2R_g$.
This defines the impact parameter of the light ray $b=r=xR_g$. At this radius the gravitational acceleration and the centrifugal acceleration are balanced but the radius of curvature of the light ray is very large compared to the orbital radius with the exception of the limiting radius of the event horizon $1R_g$ where both are equal. So light can escape the black hole for $r>1R_g$ but cannot escape for $r<1R_g$. In general if the orbiting light is moving toward the black hole center, it is likely to gain energy from its surroundings because the energy density near the center is higher. if it is moving away from the center, it will eject its energy to the surroundings. When it's energy is increased $r_\kappa$ is reduced and it is able to go into lower orbits. When it's energy is decreased $r_\kappa$ is increased and it is able to go into higher orbits. So to consider all possible curvatures of a light ray we will use a general expression for the radius of curvature given by $r_\kappa=z^2R_g$ where $z$ can have values other than $x$ and when $z=x$, it will signify balanced accelerations at $r=xR_g$. In a circular orbit, when the energy of a light ray is not constant, we have to use the de Broglie acceleration term in Eq.~\eqref{1.26a} which reduces to
\begin{align}\label{1.26b}
\left(\frac{c}{\nu}\frac{d\nu}{dt}\mathbf{\wh T}+\frac{c^2}{r_\kappa}\mathbf{\wh N}\right)=\left(\frac{\mu}{r^2}\mathbf{\wh N}\right).
\end{align}
\begin{align}\label{1.26c}
\left(\frac{c}{\nu}\frac{d\nu}{dt}\right)^2+\left(\frac{c^2}{r_\kappa}\right)^2=\left(\frac{\mu}{r^2}\right)^2.
\end{align}
\begin{align}\label{1.26d}
\left(\frac{c^2}{r_\kappa}\right)^2-\left(\frac{\mu}{r^2}\right)^2=-\left(\frac{c}{\nu}\frac{d\nu}{dt}\right)^2.
\end{align}
\begin{align}\label{1.26e}
\pm\sqrt{\left(\frac{c^2}{r_\kappa}\right)^2-\left(\frac{\mu}{r^2}\right)^2}=i\left(\frac{c}{\nu}\frac{d\nu}{dt}\right).
\end{align}
We substitute for $r=xR_g$ and $r_\kappa=z^2R_g$ and integrate. 
\begin{align}\label{1.26f}
i\int{\frac{d\nu}{\nu}}=\pm\left[\frac{c}{R_g}\sqrt{\frac{1}{z^4}-\frac{1}{x^4}}\right]\int{dt}.
\end{align}
This integration does not affect the orbital parameter $x$ but only the curvature parameter $z$. We can substitute $dt=(r/c)d\phi$ which gives for one convolution
\begin{align}\label{1.26g}
i\int^{\nu_2}_{\nu1}{\frac{d\nu}{\nu}}=\pm\left[x\sqrt{\frac{1}{z^4}-\frac{1}{x^4}}\right]\int^{2\pi}_{0}{d\phi}.
\end{align}
\begin{align}\label{1.26h}
i[\text{ln}\nu_2-\text{ln}\nu_1]=\pm 2\pi x\sqrt{\frac{1}{z^4}-\frac{1}{x^4}}.
\end{align}
This gives two complex expressions.
\begin{align}\label{1.26i}
\sqrt{\frac{1}{z^4}-\frac{1}{x^4}}=\mp i\frac{[\text{ln}\nu_2-\text{ln}\nu_1]}{2\pi x}.
\end{align}
We see that for $\nu1=\nu2$, $z=x$, Eq.~\eqref{1.26i} becomes indeterminate. This gives two solutions for $z$. Plus sign for increase in energy and minus for decrease. $m$ is introduced for more number of convolutions.
\begin{align}\label{1.26j}
z=\left[\frac{1}{x^4}-\left(\frac{\text{ln}\nu_2-\text{ln}\nu_1}{2\pi mx}\right)^2\right]^{-0.25}.
\end{align}
During EHT study of M87 BH \cite{179}, it was established that the photon ring is consistent with the strong gravitational lensing of synchrotron emission from a hot plasma orbiting near the black hole event horizon. The jet of M87 is powered by extraction of black hole spin energy through mechanisms related to the Blandford-Znajek process \cite{177}. The M87 jet is visible with similar morphology at all wavelengths from radio to X-ray. The optical radiation from the jet on kpc scales was found to be linearly polarized suggesting that the emission mechanism is synchrotron radiation. The mass of M87 black hole that powers the jet was measured from the central stellar velocity dispersion and directly from the size of the observed emitting region surrounding the black hole shadow of $5R_g$ \cite{180}. The high-angular-resolution observation with the EHT allows to reconstruct the geometry of magnetic fields in the immediate vicinity of the event horizon of the M87 supermassive black hole. The physical interpretation of polarimetric images and the discussion of horizon-scale magnetic field geometries consistent with these EHT images are presented in ref. \cite{181a}. EHT observations at $230$ GHz have imaged polarized emission around the  event-horizon of M87 BH. This polarized synchrotron radiation probes the structure of magnetic fields and the plasma properties near the black hole. It is noted that magnetic fields of $B\geq 5$ G are sufficient to produce jet powers of $P_{jet}\geq 10^{42}$ erg $s^{-1}$ via the Blandford and Znajek process \cite{177}. While pioneering attempts have been made \cite{182,183,184}, it is not yet known how to properly incorporate electron acceleration in global GRMHD simulations of hot accretion flows. However, we shall see later how periodic relativity (PR) can contribute to this area.

While actual imaging of M87 black hole's core was done at a much lower radio frequency of 230 GHz (1.3 mm wavelength), The EHTC also partnered with several international facilities in space and on the ground, to arrange an extensive, quasi-simultaneous multi-wavelength campaign \cite{185}. The Swift Observatory is equipped with UV/Optical Telescope as well as with X-ray imaging optics. Algaba et al. \cite{185} calculated the UVOT spectral index of the M87 core region during the EHT campaign and found it to be consistent with the optical/UV spectral index of the core of M87 reported by Perlman et al. \cite{186}. This article \cite{185} also proposes double zone modeling, one for the EHT-observed region and the other for the inner jets. And it concludes that VHE $\gamma$-rays do not originate from EHT-observed region. In periodic relativity (PR) we propose that VHE $\gamma$-rays can occur due to internal gravity assisted wave stacking in the proximity of the event horizon of M87. Ten hard x-rays having same phase, stacked together can produce one coherent $\gamma$-ray. The process is the same as constructing a laser beam but what keeps the component waves together is the gravitational attraction between the relativistic masses $E/c^2$ of the waves. Gravitation within electromagnetic wave is discussed in Section~\ref{gemw}.  

Spectral energy distribution (SED) for the M87 core detected by quasi-simultaneous multi-wavelength campaign confirmed presence of wavelengths ranging from 1.7 GHz to $3.82 \times 10^{26}$Hz as listed in Table A8 of ref. \cite{185}. In order to generate data from Eq.~\eqref{1.26j}, we will use light frequency of $\nu1=500$ THz at photon ring radius of $\sqrt{27}R_g$ \cite{151,152} from F606W band observed by Hubble Space Telescope (HST). A light ray cannot spiral into a black hole without an imbalance between its centrifugal acceleration and the gravitational acceleration pulling it. This imbalance comes from the slower rate of energy absorption by light ray from its surrounding compared to the faster rate of the gravitational pull. The mechanism of energy absorption is the internal gravity assisted wave stacking. We will use a typical frequency of U band observed by the Swift observatory as $\nu2=872.7$ THz which is at the lower end of the ultraviolet spectrum. Ultraviolet and higher energy rays are not visible to human eyes. As a light ray in accretion disk absorbs energy from its surroundings and spirals into the black hole we can expect deviation of $z$ from $x=\sqrt{27}=5.196$ which we can compute for $m=1$ convolution using Eq.~\eqref{1.26j}. It comes out to be $z=5.515$. If we check for $x=6.1$ with same frequencies, we get $z=6.651$. Value of $z=5.515$ match the measured values of the shadow of M87 which is $10.4\leq \alpha \leq 11.5$. This shows that the shadow of a black hole is due to frequency shift of light into ultraviolet and higher range at $\approx5R_g$. Eq.~\eqref{1.26j} is not dependent on the mass or the spin of a black hole. Therefore same explanation is applicable to SgrA* \cite{173}. Data shown in Table~\ref{tab:Table1} using Eq.~\eqref{1.26j} provides the proof that the unaltered synchrotron radiation cannot exist for $z\leq 1.8$. $1.8 R_g$ is the radius at which gravity assisted wave stacking begins to create $\gamma$-rays. These stacked $\gamma$-rays can escape as jets from the poles because in PR the event horizon is at $1R_g$ when total mass (including spin energy) is used for defining $R_g$ and the event horizon is between $1R_g$ and $\sqrt{2}R_g$ when irreducible mass is used for defining $R_g$.

\begin{table}
	 \caption{Deviation of $z$ due to light ray absorbing energy.\label{tab:TableA}}
 \begin{ruledtabular}
		\begin{tabular} {llllll}
		$x$ &$\nu_1$THz &$\nu_2$THz &$m$ &$\phi$\:rad &$z$\\  \hline 
    $6.1$ &$500$\,yellow  &$872.7$\,$ultraviolet$ &$1$ &$2\pi$   &$6.651$\\
		$5.196$ &$500$\,yellow  &$872.7$\,$ultraviolet$ &$1$ &$2\pi$   &$5.515$\\
		$5.196$ &$500$\,yellow	&$2E5$\,x-ray soft &$1$ &$2\pi$ &$2.359$\\
		$5.196$ &$500$\,yellow &$3.8E6$\,x-ray hard &$1$ &$2\pi$ &$1.92$\\
		$5.196$ &$500$\,yellow  &$1.154E7$\,$\gamma$-ray &$1$ &$2\pi$ &$1.809$\\
		$5.196$ &$500$\,yellow  &$6E8$\,$\gamma$-ray &$1$ &$2\pi$ &$1.530$\\
		$5.196$ &$500$\,yellow  &$6E8$\,$\gamma$-ray &$0.5$ &$\pi$ &$1.0805$\\
		$5.196$ &$500$\,yellow  &$6E8$\,$\gamma$-ray &$0.42852$ &$2.69247$ &$1.000$\\
    $5.196$ &$500$\,yellow  &$2.4E23$\,Cosmic ray &$1$ &$2\pi$ &$0.8282$\\
    $5.196$ &$500$\,yellow  &$2.4E23$\,Cosmic ray &$0.1$ &$0.6283$ &$0.2618$\\
    $5.196$ &$500$\,yellow  &$2.4E23$\,Cosmic ray &$0.01$ &$0.0628$ &$0.0827$\\
				\end{tabular}
 \end{ruledtabular}
\end{table}
Selection of value of $m$ is ad hoc. For data corresponding to $m=0.43$ in Table~\ref{tab:Table1}, we can see that the yellow light ray taking a plunge at the equatorial plane having the accretion disk can approach the event horizon $z=x=1$ as a $\gamma$-ray and without entering the event horizon can exit from two poles of the black hole forming $\gamma$-ray jets. Blandford and Znajek \cite{177} have discussed circumstances under which the energy and angular momentum of a rotating black hole can be extracted electromagnetically. If the field lines have paraboloidal shape, the energy is beamed along the rotation axis.  

\subsubsection{Inside the event horizon} \label{Inside}
Last three data sets in Table~\ref{tab:Table1} deals with the Cosmic rays. The maximum energy of cosmic ray recorded to date is of the order of $10^{20}eV$ which corresponds to a frequency of $2.4\times10^{23}$THz. This gives value of $z=x=0.8$ to $0.08$. $m=0.0628$ is almost a radial motion and originates at the center of the $1R_g$ radius of the event horizon. And the equations of PR do not fail here. Eq.~\eqref{1.26j} keeps generating similar data with frequencies up to $10^{308}$Hz. Beyond that I get computer overflow but the equations don't fail. If the mass of M87 BH, which is $1.29E+40$ Kg can be contained in a massless pair of particle anti-particle following the quantum condition $E=h\nu$, then the frequency of those particles will be of the order $\nu=8.95E89$\,Hz. This puts a limit on the maximum frequency at the center of the M87 BH. The energy of the M87 BH manifest to the human senses is only a small fraction of this energy. From above discussion it is clear that there may not be any light orbiting at $1R_g$ radius of the event horizon, but only x-rays and $\gamma$-rays. Therefore the current EH Telescope technology may not be adequate for probing the event horizon of a black hole.

So what could be the nature of the singularity that existed prior to the big bang? GR cannot go beyond $10^{-43}$\,sec.$\equiv 10^{43}$\,Hz. The estimated energy of the entire universe including dark matter and dark energy is about $1.823E81$\,GeV ($2.92E71$\,J.) This converted to the frequency of a massless pair of particles comes to about $\nu=2.2E104$\,Hz. This is a small value compared to what PR can handle.

No matter what values of frequencies are used in Eq.~\eqref{1.26j}, the parameter $z$ remains non-zero as long as $x\neq 0$. The value of $x$ becomes zero only when the two gravitationally bound objects become one single object. This is when $x=z=0$. When this concept is extended to the energies of the entire universe, we can see that there exists a transcendental non-dual state of energy which is perfectly motionless and without any divisions. This is the fundamental substance (or rather field) of the universe from which the entire universe is created.

From above discussion, it becomes clear that Fermions cannot satisfy the condition $x=z=0$, but only Bosons can do so. So there is a good reason to believe that all the fermions are converted to bosons with in the ergosphere of a black hole and within the event horizon and most of the content within the event horizon must be superfluid boson condensates. As long as boson condensates maintain their separate identity from the underlying motionless indivisible field of the fundamental substance, the final union has not yet taken place. When the boson condensates begin to disappear from the universe, that would be the sign that the final union has taken place. However, it is possible that the universe can bounce back from the state of superfluid boson condensates and that can happen for several cycles prior to final dissolution.

In a bounce back scenario, the superfluid boson condensates would first disintegrate into cosmic rays and when they cross the event horizon, the result would be a $\gamma$-ray burst. This is the reversal of the gravity assisted wave stacking process. Published version of Section~\ref{BH} is available in \cite{222}. 

\subsubsection{Electron acceleration in global PRMHD simulations of hot accretion flows} 
Electron acceleration in an electromagnetic field surrounding black hole event horizon can be modeled in PR using the expression for Lorentz invariant acceleration used in Eq.~\eqref{1.26} as follows.
\begin{align}\label{1.26k}
\begin{split}
\mathbf{a}_l&=\left(\left(\frac{qE}{m_e}+\frac{v}{\nu}\frac{d\nu}{dt}\right)\mathbf{\wh T}+\left(\frac{qvB}{m_e}+\kappa v^2\right)\mathbf{\wh N}\right)\\&=
\left(0\mathbf{\wh T}+\frac{\mu}{r^2}\left(\cos{(\psi-\gamma)}+\sin{(\psi-\gamma)}\right)\mathbf{\wh N}\right),
\end{split}
\end{align}
where $q$ is electron charge, $E$ Electric field strength, $m_e$ electron mass, $v$ electron velocity, $\nu$ electron frequency, $B$ magnetic field strength, $\kappa$ curvature of electron trajectory (paraboloid). Therefore the deviation factor $n$ for the PR line element Eq.~\eqref{2.2m} can be written as 
\begin{align}\label{1.26l}
n=\frac{\left((qE/m_e)+(v/\nu)(d\nu/dt)\mathbf{\wh T}+((qvB/m_e)+\kappa v^2)\mathbf{\wh N}\right)}
{\left(0\mathbf{\wh T}+(\mu/r^2)\left(\cos{(\psi-\gamma)}+\sin{(\psi-\gamma)}\right)\mathbf{\wh N}\right)}=1.
\end{align} 

\section{Einstein's field equations}\label{Efe}
\hspace*{5 mm} Now we are in a position to write Eq.~\eqref{1.18} in a metric form as follows.
\begin{align}\label{4.40}
ds^2=g_{\mu\nu} dx^\mu dx^\nu.
\end{align}
\begin{align}\label{4.44}
g_{\mu\nu}=
\begin{pmatrix}
	c^2 & 0 & 0 & 0 \\
	0 & -n & 0 & 0\\
	0 & 0 & -nr^2 & 0\\
	0 & 0 & 0 & -n(r^2\sin^2{\theta})
\end{pmatrix}.
\end{align}
We can adjust the value of deviation factor $n$ to match the observed value of the perihelic precession for individual planets. As a matter of fact all future strong field variations in general relativity could be explained by adjusting this parameter $n$. This kind of adjustment is not possible in general relativity and other metric theories because that will affect the predicted values of deflection of light, gravitational redshift and the limiting radius of event horizon. This factor $n$ may have an internal structure dependent on the natural frequency and composition of the orbiting body (Doppler frequency shift of the constituent massive particles of the body). If we alter this factor $n$ in Eq.~\eqref{1.17i} with any suitable constant then it will always satisfy Einstein's field equations.\\
\hspace*{5 mm}Here we verify to see that Eq.~\eqref{1.18} does satisfy Einstein's field equations. For this purpose it will be necessary to calculate Christoffel symbols $\Gamma_{\mu \nu}^\sigma$. At the same time the proper time interval should be experimentally verified because all deviations and variations get accumulated in the expression for proper time and any error in the theory would show up there as well.\\
\hspace*{5 mm}The metric ~\eqref{4.44} is diagonal, so the non-zero components of the contravariant metric tensor are $g^{\sigma\sigma} = 1/g_{\sigma\sigma}$. Hence the diagonality of the metric allows us to simplify the definition of the Christoffel symbols to
\begin{align}\label{4.45}
\Gamma_{\mu \nu}^\sigma=\frac{1}{2}g^{\sigma\sigma}\left(\frac{\partial g_{\sigma\mu}}{\partial x^\nu}+ \frac{\partial g_{\sigma\nu}}{\partial x^\mu}-\frac{\partial g_{\mu\nu}}{\partial x^\sigma}\right),
\end{align}
where the suffixes assume four values 0, 1, 2, 3 and no summations are implied. We consider the case of static spherically symmetric field produced by a spherically symmetric body at rest. Line element given by Eq.~\eqref{1.18} is compatible with spherical symmetry. Coordinate $x^0$ is taken to be time $t$, and the spatial coordinates may be taken to be spherical polar coordinates $x^1=r$, $x^2=\theta$, $x^3=\phi$. We can determine the values of $g_{\mu\nu}$ from metric ~\eqref{4.44},
\begin{align}\label{4.46}
\begin{split}
&g_{00}=c^2, \quad g_{11}=-n, \quad g_{22}=-nr^2, \quad g_{33}=-nr^2\sin^2{\theta},\\ &g^{\mu\nu} = 1/g_{\mu\nu} \quad and \quad g_{\mu\nu}=0, \quad g^{\mu\nu}=0 \quad for \quad \mu\neq\nu.
\end{split}
\end{align}
Inserting these values into Eq.~\eqref{4.45} we find that the only non-vanishing Christoffel symbols are 
\begin{align}\label{4.47}
\begin{split}
&\Gamma_{11}^1=\frac{1}{2n}\frac{\partial n}{\partial r} \qquad \qquad \qquad \hspace*{2 mm} \Gamma_{33}^2=-\sin{\theta}\cos{\theta}\\
&\Gamma_{22}^1=-r-\frac{r^2}{2n}\frac{\partial n}{\partial r}\qquad \qquad \Gamma_{13}^3=\Gamma_{31}^3=\frac{1}{r}+\frac{1}{2n}\frac{\partial n}{\partial r}\\
&\Gamma_{33}^1=-r\sin^2{\theta}\left(1+\frac{r}{2n}\frac{\partial n}{\partial r}\right) \qquad \Gamma_{23}^3=\Gamma_{32}^3=\cot{\theta}.\\
&\Gamma_{12}^2=\Gamma_{21}^2=\frac{1}{r}+\frac{1}{2n}\frac{\partial n}{\partial r}
\end{split}
\end{align}
The expression for the Ricci tensor is
\begin{align}\label{4.48}
R_{\mu \nu}=\Gamma_{\mu\alpha,\:\nu}^\alpha-\Gamma_{\mu\nu,\:\alpha}^\alpha-\Gamma_{\mu\nu}^\alpha\Gamma_{\alpha\beta}^\beta+\Gamma_{\mu\beta}^\alpha\Gamma_{\nu\alpha}^\beta.
\end{align}
Einstein's law of gravitation requires the Ricci tensor to vanish $(R_{\mu \nu}=0)$ in empty space. We can now write the components of the Ricci tensor, each of which must vanish in order for the field equations to be satisfied. From symmetry arguments we can expect all the non-diagonal components to be zero. Hence the only components of interest in case of our line element are the diagonal elements. 
Substitution of Eq.~\eqref{4.47} in Eq.~\eqref{4.48} gives 
\begin{align}
&R_{00}=0,\label{4.53}\\
&R_{11}=\frac{1}{nr}\frac{\partial n}{\partial r}-\frac{1}{n^2}\left(\frac{\partial n}{\partial r}\right)^2+\frac{1}{n}\frac{\partial^2 n}{\partial r^2},\label{4.54}\\
&R_{22}=\frac{3r}{2n}\frac{\partial n}{\partial r}-\frac{r^2}{4n^2}\left(\frac{\partial n}{\partial r}\right)^2+\frac{r^2}{2n}\frac{\partial^2 n}{\partial r^2},\label{4.55}\\
&R_{33}=R_{22}\sin^2{\theta}.\label{4.56}
\end{align}
The vanishing of Eq.~\eqref{4.54} leads to
\begin{align}\label{4.57}
\frac{\partial^2 n}{\partial r^2}=\frac{1}{n}\left(\frac{\partial n}{\partial r}\right)^2-\frac{1}{r}\frac{\partial n}{\partial r}.
\end{align}
Substituting of Eq.~\eqref{4.57} in Eq.~\eqref{4.55} and equating it to zero gives the condition for vanishing of the Ricci tensor Eq.~\eqref{4.48}.
\begin{align}\label{4.58}
\left(\frac{r}{n}\frac{\partial n}{\partial r}\right)^2+4\left(\frac{r}{n}\frac{\partial n}{\partial r}\right)=0.
\end{align}
This quadratic equation has two solutions.
\begin{align}\label{4.58a}
\left(\frac{r}{n}\frac{\partial n}{\partial r}\right)=0 \qquad and \qquad \left(\frac{r}{n}\frac{\partial n}{\partial r}\right)=-4.
\end{align}
This shows that any constant value of $n$ will satisfy the first solution. This means that our derivation of gravitational redshift, deflection of light, perihelic precession of planets and the limiting radius of event horizon of a black hole are exact solutions of Einstein's field equations. These solutions however are at variance with the Schwarzschild solution.

\section{Rotation curves of galaxies}\label{RC}
In Section~\ref{Efe} we obtained two solutions to Einstein's field equations,
\begin{align}\label{4.58am}
\left(\frac{r}{n}\frac{\partial n}{\partial r}\right)=0 \qquad and \qquad \left(\frac{r}{n}\frac{\partial n}{\partial r}\right)=-4.
\end{align}
So far we have seen the application of the first solution which requires $n$ to be any real number constant. Now we look at the second solution which we can write as
\begin{align}\label{1.31m}
\int\frac{\partial n}{n}=-4\int\frac{\partial r}{r}.
\end{align}
\begin{align}\label{1.32m}
\text{ln}(nr^4)=C.
\end{align}
where $C$ is a constant of integration. This gives
\begin{align}\label{1.33}
n=\frac{e^C}{r^4}=\frac{k}{r^4}.
\end{align}
\begin{align}\label{1.34}
\text{where}\ \ \  k=e^C=\text{constant}.
\end{align}
In this second solution $n$ need not be a constant. Please note that an improved published version of Section~\ref{RC} based on first solution to Einstein's field equations is available in \cite{222}. Here we make use of Eq.~\eqref{1.26} in order to apply the second solution to rotation curves of a galaxies. Assuming circular orbit we substitute $\psi=\pi/2$ and $\gamma=0$. This gives
\begin{align}\label{1.35m}
|\mathbf{a}|=\kappa\left(\frac{ds}{dt}\right)^2=\frac{v^2}{r}=\frac{\mu}{r^2}.
\end{align}
Then we define unitless deviation factor $n$ as a ratio of Newtonian acceleration to the observed acceleration given by
\begin{align}\label{1.39}
n=\frac{\mu/r^2}{v^2/r}.
\end{align}
This ratio is $n=1$ for circular orbits in flat Minkowski space time. Equating Eq.~\eqref{1.39} with Eq.~\eqref{1.33} we get,
\begin{align}\label{1.40}
n=\frac{\mu}{v^2r}=\frac{k}{r^4}.
\end{align}
This gives,
\begin{align}\label{1.36}
k=\frac{\mu r^3}{v^2}.
\end{align}
Value of $k$ given by Eq.~\eqref{1.36} is a constant of orbit. This means that every star orbit in a galaxy will have its own constant $k$. 
We can write Eq.~\eqref{1.40} as
\begin{align}\label{1.37}
v^2=\frac{4\pi^2r^2}{P^2}=\frac{\mu}{nr}.
\end{align}
\begin{align}\label{1.37a}
P=\frac{2 \pi r}{v}.
\end{align}
\begin{align}\label{1.38m}
P^2=\frac{4\pi^2r^3n}{\mu}.
\end{align}
For $n=1$, Eq.~\eqref{1.38m} reduces to Kepler's third law, where $P$ is the orbital period.  
By substituting Eq.~\eqref{1.40} in Eq.~\eqref{1.17} we can compute the ratio $d\tau/dt$. We can apply these equations of stellar motion to Blue Horizontal-Branch (BHB) halo stars of the Milky Way \cite{191}. The circular velocity estimates are based on Naab's simulation \cite{41}. To this data, one additional data point for solar radius of $8kpc$ \cite{26} is added and the results obtained from Eqs.~\eqref{1.36},~\eqref{1.33} and~\eqref{1.17} are shown in Table~\ref{tab:Table1}. Computed values are based on the stellar mass at the galactic center, which is $5.0924\times10^{10}M_\odot$ \cite{181,42}. Observed values of $r$ and circular velocities constrain the integration constant $k$ which provides a measure of non-uniform distribution of the galactic matter and the cold dark matter at a given radius. Hence it is appropriate to describe $k$ as a galactic matter distribution constant. We also find that Eqs.~\eqref{1.37a} and~\eqref{1.38m} both yield exactly the same orbital period when velocity and deviation $n$ along with the galactic stellar mass are used from the Tables. For the Sun, both yield 223.4 million years.
\begin{table}
	 \caption{Milky Way rotation curve based on proper time.\label{tab:Table1}}
 \begin{ruledtabular}
		\begin{tabular} {llllll}
		$r (kpc)$ &$v (km/s)$ &$k\times 10^{-81}$ &$\hspace*{7 mm}n$ &$d\tau/dt$ \\  \hline 
		$7.5$ &$216$ &$1.79546$ &$0.62593$ &$1-1.6246\times 10^{-7}$\\
		$8.0$ &$220$	&$2.10050$ &$0.56566$ &$1-1.5231\times 10^{-7}$\\
		$12.5$ &$227$ &$7,52624$ &$0.34004$ &$1-9.748\times 10^{-8}$\\
		$17.5$ &$179$  &$33.2129$ &$0.39061$ &$1-6.9628\times 10^{-8}$\\
		$22.5$ &$168$ &$80.1362$ &$0.34490$ &$1-5.4155\times 10^{-8}$\\
		$27.5$ &$183$ &$123.309$ &$0.23782$ &$1-4.43091\times 10^{-8}$\\
		$32.5$ &$143$ &$333.332$ &$0.32956$ &$1-3.7492\times 10^{-8}$\\
		$37.5$ &$170$ &$362.322$ &$0.20210$ &$1-3.2493\times 10^{-8}$\\
		$42.5$ &$183$ &$455.160$ &$0.15388$ &$1-2.8670\times 10^{-8}$\\
		$47.5$ &$165$ &$781.650$ &$0.16936$ &$1-2.5652\times 10^{-8}$\\
		$55$ &$183$ &$986.474$ &$0.11891$ &$1-2.2154\times 10^{-8}$\\
				\end{tabular}
 \end{ruledtabular}
\end{table}

\begin{table}
	 \caption{Solar system rotation curve based on proper time.\label{tab:Table2}}
 \begin{ruledtabular}
		\begin{tabular} {llllll}
Planet	&$r\times10^{-9}(m)$ &$v (km/s)$ &$k$ &$\hspace*{3 mm}n$ \\  \hline 
Mercury	&$57.91$ &$47.87$ &$1.12\times 10^{43}$ &$1.000103$ \\
Venus		&$108.21$ &$35.02$	&$1.37\times 10^{44}$ &$1.000059$ \\
Earth		&$149.6$ &$29.78$ &$5.01\times 10^{44}$ &$1.000332$ \\
Mars		&$227.92$ &$24.13$  &$2.69\times 10^{45}$ &$1.000065$ \\
Jupiter	&$778.57$ &$13.07$ &$3.66\times 10^{47}$ &$0.997876$ \\
Saturn	&$1433.53$ &$9.69$ &$4.16\times 10^{48}$ &$0.985986$ \\
Uranus	&$2872.46$ &$6.81$ &$6.78\times 10^{49}$ &$0.99627$ \\
Neptune	&$4495.06$ &$5.43$ &$4.08\times 10^{50}$ &$1.00136$ \\
Pluto		&$5869.66$ &$4.72$ &$1.20\times 10^{51}$ &$1.014912$ \\
Moon		&$0.3844$ &$1.023$ &$2.16\times 10^{34}$ &$0.990824$ \\
				\end{tabular}
 \end{ruledtabular}
\end{table}

\begin{table}
	 \caption{M31 rotation curve. $k$ in m$^4$, $P$ in yrs.\label{tab:Table3}}
 \begin{ruledtabular}
		\begin{tabular} {llllll}
		$r (kpc)$ &$v (km/s)$ &$k\times 10^{-81}$ &$\hspace*{1 mm}n$ &$d\tau/dt$ &$P\times 10^{-8}$ \\  \hline 
		$8.5$ &$232.0$ &$6.23$ &$1.316$ &$1-3.94\times 10^{-7}$ &$2.250$\\
		$12.5$ &$251.2$	&$16.89$ &$0.763$ &$1-2.68\times 10^{-7}$ &$3.057$\\
		$16.5$ &$251.6$ &$38.74$ &$0.576$ &$1-2.03\times 10^{-7}$ &$4.029$\\
		$20.5$ &$227.4$  &$90.94$ &$0.568$ &$1-1.63\times 10^{-7}$ &$5.538$\\
		$24.5$ &$226.2$ &$156.89$ &$0.480$ &$1-1.367\times 10^{-7}$ &$6.654$\\
		$28.5$ &$218.8$ &$263.96$ &$0.441$ &$1-1.175\times 10^{-7}$ &$8.0$\\
		$32.5$ &$224.7$ &$371.15$ &$0.367$ &$1-1.030\times 10^{-7}$ &$8.885$\\
		$36.5$ &$240.1$ &$460.47$ &$0.286$ &$1-9.178\times 10^{-8}$ &$9.339$\\
				\end{tabular}
 \end{ruledtabular}
\end{table}

\begin{table}
	 \caption{NGC3198 rotation curve. $k$ in m$^4$, $P$ in yrs.\label{tab:Table4}}
 \begin{ruledtabular}
		\begin{tabular} {llllll}
$r (kpc)$ &$v (km/s)$ &$k\times 10^{-79}$ &$\hspace*{1 mm}n$ &$d\tau/dt$ &$P\times 10^{-8}$ \\  \hline 
		$0.68$ &$55$ &$0.202$ &$10.45$ &$1-1.76\times 10^{-7}$ &$0.759$\\
		$1.36$ &$92$	&$0.579$ &$1.868$ &$1-8.79\times 10^{-8}$ &$0.908$\\
		$2.72$ &$123$ &$2.593$ &$0.522$ &$1-4.39\times 10^{-8}$ &$1.358$\\
		$5.44$ &$147$  &$14.52$ &$0.183$ &$1-2.2\times 10^{-8}$ &$2.273$\\
		$8.16$ &$156$ &$43.52$ &$0.108$ &$1-1.466\times 10^{-8}$ &$3.213$\\
		$13.6$ &$154$ &$206.78$ &$0.066$ &$1-8.79\times 10^{-9}$ &$5.425$\\
		$19.04$ &$148$ &$614.36$ &$0.0515$ &$1-6.28\times 10^{-9}$ &$7.903$\\
		$24.48$ &$148$ &$1305.7$ &$0.040$ &$1-4.88\times 10^{-9}$ &$10.16$\\
		$29.92$ &$149$ &$2352.1$ &$0.0323$ &$1-3.99\times 10^{-9}$ &$12.33$\\
				\end{tabular}
 \end{ruledtabular}
\end{table}

Table~\ref{tab:Table2} shows solar system data from NASA planet fact sheets. Radial distance equal to semi major axis and mean orbital velocity are used. $k$ and $n$ are computed using Eqs.~\eqref{1.36} and~\eqref{1.33}. $(1-d\tau/dt)$ are of order $10^{-8}$ to $10^{-12}$ and not shown in the table. In case of moon, earth mass $5.9736\times10^{24}$ Kg. is used. Value of $n$ for Mercury shown in Table~\ref{tab:Table2} should not be compared with that used in the Eq.~\eqref{1.17i} for the perihelic precession of planets because here we have used second solution of Einstein's field equations with constant $k$, where as perihelic precession is derived from the first solution of Einstein's field equations with constant $n$. These two solutions are derived from two roots of a quadratic equation. The purpose of presenting the solar system data is only to show that there is no discontinuity like the MOND function. One should not look for precision in Table~\ref{tab:Table2} because it is based on circular orbit approximation. It is sufficient to note that $n=1$ for flat Minkowski metric is recovered at small distances.

We can also apply these equations of stellar motion to rotation curves of M31 \cite{52} and NGC3198 \cite{53}. The results obtained from Eqs.~\eqref{1.36},~\eqref{1.33} and~\eqref{1.17} are shown in Tables~\ref{tab:Table3} and~\ref{tab:Table4}. Computed values are based on the stellar mass at the galactic center, which is $1.4\times10^{11}M_\odot$ for M31 and $5.0\times10^{9}M_\odot$ for NGC3198.  

\section{Perihelic precession of electron in hydrogen atom}
Periodic relativity is a theory of accelerations. Whether the acceleration is due to gravity or Coulomb force, it does not matter. In case of rotation curves of galaxies, we defined deviation factor $n$ as a unitless ratio of accelerations given by Eq.~\eqref{1.39}. We used the same method to define the ratio of Coulomb acceleration and radial acceleration of electron to obtained energy levels of hydrogen atom \cite{153} without the use of potential energy term of the Schrodinger and Dirac wave equations. Here we apply the same formalism of Section~\ref{ppp} for perihelic precession of planets to analyze the perihelic precession of electron in hydrogen atom. 

We have defined $\bar{n}$ for Hydrogen atom as the ratio of acceleration due to Coulomb force and the centrifugal  acceleration which is indirectly observed in the form of spectra \cite{153}.
\begin{align}\label{1.35b}
\bar{n}=\frac{-e^2kZ/m_0r^2}{v^2/r}=\frac{-e^2kZ}{m_0v^2r},
\end{align}
where $e$ is the electric charge, $m_0$ the electron rest mass, $k$ Coulomb's constant, $Z$ atomic number (1 for Hydrogen), $v$ is the orbital velocity of electron, and $r$ the radial distance. We also discovered that in our relativistic formalism the orbital velocity of electron to the first order accuracy is about $30\%$ less than that in the Bohr model and is given by,
\begin{align}\label{1.27m}
v\approx\left(\frac{e^2kZ}{\sqrt{2}\;\hbar n}\right),
\end{align}
where $\hbar$ is the Planck constant and $n$ the principal quantum number. If we substitute Eq.~\eqref{1.27m} and $r=a_0n^2$ in Eq.~\eqref{1.35b}, we get $\bar{n}=2$. But if we substitute Bohr velocity then we get $\bar{n}=1$. For this reason we cannot start our formalism like Eq.~\eqref{1.16a}, but we have to start from fundamentals by first establishing the principle of equivalence as applicable to hydrogen atom. Equivalence of gravitational mass and inertial mass in Einstein's theory follows from equivalence of gravitational acceleration and inertial acceleration. In a specific case of orbital motion of planets, we can restrict this definition of equivalence of accelerations to following statement.
\begin{itemize}
	\item In Keplerian circular orbit, the gravitational acceleration is equal and opposite to the centrifugal acceleration acting on the body.
\end{itemize}
Newton started out with this equality and introduced Kepler's third law of orbital periods in it and thus arrived at the inverse square law of gravitation. From this we can define the principle of equivalence as applicable to Hydrogen atom.
\begin{itemize}
	\item In Keplerian circular orbit, the electromagnetic acceleration is equal and opposite to the centrifugal acceleration acting on the body.
\end{itemize}
This definition of equivalence of accelerations is not possible in Schrodinger or Dirac theory because they have abandoned the velocity parameter. Thus we can write, 
\begin{align}\label{1.16aa}
\frac{d^2\mathbf{r}}{dt^2}=-\frac{v^2}{r}\mathbf{\hat r}=
-\frac{ke^2Z}{2m_0r^2}\mathbf{\hat r}.
\end{align}
\begin{align}\label{1.16ab}
m_0\frac{d^2\mathbf{r}}{dt^2}=-\frac{ke^2Z}{2r^2}\mathbf{\hat r}.
\end{align}
Eq.~\eqref{1.16ab} gives inverse square law of the electromagnetic attraction which is not same as the Coulomb's law of electrostatic attraction. Now we introduce the line element of periodic relativity Eq.~\eqref{1.31} in Eq.~\eqref{1.16aa}. With this we replace the coordinate time interval $dt$ of Eq.~\eqref{1.16aa} with the proper time interval $d\tau$. 
\begin{align}\label{1.32mm}
\frac{d^2\mathbf{r}}{d\tau^2}(1-\bar{n}\beta^2)=-\frac{ke^2Z}{2m_0r^2}\mathbf{\hat r}.
\end{align} 
\begin{align}\label{1.32am}
\frac{d^2\mathbf{r}}{d\tau^2}=-\left(\frac{ke^2Z}{2m_0r^2}\right)(1-\bar{n}\beta^2)^{-1}\mathbf{\hat r}.
\end{align} 
\begin{align}\label{1.32bm}
\frac{d^2\mathbf{r}}{d\tau^2}=-\left(\frac{ke^2Z}{2m_0r^2}\right)(1+\bar{n}\beta^2)\mathbf{\hat r}.
\end{align} 
We will write $v^2$ on r.h.s. in polar coordinates. 
\begin{align}\label{1.17cm}
\begin{split}
&\frac{d^2\mathbf{r}}{d\tau^2}=-\left(\frac{ke^2Z}{2m_0r^2}\right)\mathbf{\hat r}-\left(\frac{\bar{n}ke^2Z}{2m_0r^2c^2}\right) \left[\left(\frac{dr}{dt}\right)^2+r^2\left(\frac{d\theta}{dt}\right)^2+r^2 \sin^2\theta \left(\frac{d\phi}{dt}\right)^2 \right]\mathbf{\hat r}.
\end{split}
\end{align}
Working in $(r,\theta)$ plane we can put $d\phi=0$ and on r.h.s. substitute 
\begin{align}\label{1.18am}
r^2\left(\frac{d\theta}{dt}\right)^2=\frac{H^2}{r^2}.
\end{align}
Hence Eq.~\eqref{1.17cm} reduces to
\begin{align}\label{1.17dm}
\frac{d^2\mathbf{r}}{d\tau^2}=-\left(\frac{ke^2Z}{2m_0r^2}\right)\mathbf{\hat r}-\left(\frac{\bar{n}ke^2Z}{2m_0r^2c^2}\right)
\left[\left(\frac{dr}{dt}\right)^2+\frac{H^2}{r^2} \right]\mathbf{\hat r}.
\end{align}
On l.h.s. we substitute
\begin{align}\label{4.18m}
\frac{d^2\mathbf{r}}{d\tau^2}=\left(\frac{d^2 r}{d\tau^2}-\frac{H^2}{r^3}\right)\mathbf{\hat r}.
\end{align}
\begin{align}\label{1.17em}
\begin{split}
\left(\frac{d^2 r}{d\tau^2}-\frac{H^2}{r^3}\right) =&-\left(\frac{ke^2Z}{2m_0r^2}\right)-\left(\frac{\bar{n}ke^2Z}{2m_0r^2c^2}\right)
\left(\frac{dr}{dt}\right)^2
-\left(\frac{\bar{n}ke^2ZH^2}{2m_0r^4c^2}\right).
\end{split}
\end{align}
\begin{align}\label{1.17fm}
\begin{split}
\frac{d^2 r}{d\tau^2}=&-\left(\frac{ke^2Z}{2m_0r^2}\right)-\left(\frac{\bar{n}ke^2Z}{2m_0r^2c^2}\right)
\left(\frac{dr}{dt}\right)^2
-\left(\frac{\bar{n}ke^2ZH^2}{2m_0r^4c^2}\right)+\frac{H^2}{r^3}.
\end{split}
\end{align}
On the l.h.s. of Eq.~\eqref{1.17fm} we have the proper time and on the r.h.s. we have the coordinate time. To obtain the time independent solution of the equation of motion we make the following substitution which gives a second order non-homogeneous, non-linear differential equation.
\begin{align}\label{1.17g1m}
u=\frac{1}{r}, \quad \text{and} \quad \frac{d}{d\tau}=\frac{d}{dt}=Hu^2\frac{d}{d\theta}.
\end{align} 
\begin{align}\label{1.17gm}
\begin{split}
-H^2u^2\frac{d^2 u}{d\theta^2}=&-\left(\frac{ke^2Z}{2m_0}\right)u^2-\left(\frac{\bar{n}ke^2ZH^2}{2m_0c^2}\right)u^4+H^2u^3-\left(\frac{\bar{n}ke^2Z}{2m_0c^2}\right)u^2\left(-u\frac{du}{d\theta}\right)^2.
\end{split}
\end{align}
\begin{align}\label{1.17hm}
\begin{split}
\frac{d^2 u}{d\theta^2}+u=&\left(\frac{ke^2Z}{2m_0H^2}\right)+\left(\frac{\bar{n}ke^2Z}{2m_0c^2}\right)u^2+\left(\frac{\bar{n}ke^2Z}{2m_0c^2H^2}\right)\left(-u\frac{du}{d\theta}\right)^2.
\end{split}
\end{align}
Using the argument of Eq.~\eqref{4.40h} we can substitute,
\begin{align}\label{4.40hm}
\left(\frac{du}{d\theta}\right)^2=0.
\end{align} 
This gives,
\begin{align}\label{1.17im}
\frac{d^2 u}{d\theta^2}+u=\left(\frac{ke^2Z}{2m_0H^2}\right)+\left(\frac{\bar{n}ke^2Z}{2m_0c^2}\right)u^2.
\end{align}
We define the constant,
\begin{align}\label{1.17ima}
\eta=\left(\frac{ke^2Z}{2m_0}\right).
\end{align}
\begin{align}\label{1.17imb}
\frac{d^2 u}{d\theta^2}+u=\left(\frac{\eta}{H^2}\right)+\left(\frac{\bar{n}\eta}{c^2}\right)u^2.
\end{align}
Eq.~\eqref{1.17imb} is similar to Eq.~\eqref{1.17i} and can be solved as in general relativity. Here we have $\bar{n}=2$. Solution of Eq.~\eqref{1.17imb} in the non relativistic limit can be obtained by putting $\bar{n}=0$ and is given by 
\begin{align}\label{1.17imc}
u^{(0)}=\frac{\eta}{H^2}[1+\epsilon \cos(\theta-\theta_0)].
\end{align}
We can identify $H^2/\eta$ with the semi latus rectum of the ellipse and $\epsilon$ with the eccentricity.
For circular orbits Eq.~\eqref{1.17imc} reduces to classical Bohr orbits.
\begin{align}\label{1.17imc1}
u^{(0)}=\frac{1}{r}=\frac{1}{a_0n^2}.
\end{align}
Substituting $u^{(0)}$ back into the last term of Eq.~\eqref{1.17imb} gives an approximate differential equation which can be solved to give following approximate relativistic solution with terms significant to perihelic precession.
\begin{align}\label{1.17imd}
u=\frac{\eta}{H^2}[1+\epsilon \cos(\theta-\theta_0)]+\left(\frac{\bar{n}\eta \epsilon}{c^2}\right)\left(\frac{\eta^2}{H^4}\right)\theta \sin(\theta-\theta_0).
\end{align}
By ignoring higher order terms in $c$, Eq.~\eqref{1.17imd} becomes,
\begin{align}\label{1.17ime}
u=\frac{\eta}{H^2}[1+\epsilon \cos(\theta-\theta_0-k\theta)]
\end{align}
where,
\begin{align}\label{1.17imf}
k\theta=\left(\frac{\bar{n}\eta \theta}{c^2}\right)\left(\frac{\eta}{H^2}\right)=\left(\frac{\bar{n}\theta \eta^2 }{c^2H^2}\right).
\end{align}
Therefore in relativity, the electron orbit undergoes angular precession given by Eq.~\eqref{1.17imf}. This value of  angular precession for $\bar{n}=2$, $\theta=2\pi$ and $Z=1$ is given per orbital period $T$ by,
\begin{align}\label{1.17img}
\Delta\omega=\left(\frac{4\pi}{c^2a(1-\epsilon^2)T}\right)\left(\frac{ke^2}{2m_0}\right).
\end{align}
For near circular orbit, we can put $\epsilon=0$ and $a=a_0n^2$ which gives,
\begin{align}\label{1.17imh}
\Delta\omega=\left(\frac{2\pi}{c^2a_0n^2T}\right)\left(\frac{ke^2}{m_0}\right)=\left(\frac{2\pi\alpha^2}{n^2T}\right),
\end{align}
where $\alpha$ is the fine structure constant.

\subsection{Orbital period of electron in hydrogen atom}
We use Eq.~\eqref{1.37a} to derive the orbital period of electron in the hydrogen atom. Substitution of velocity from Eq.~\eqref{1.27m} and Bohr orbits from Eq.~\eqref{1.17imc1} gives,
\begin{align}\label{1.17imi}
P=\left(\frac{2\sqrt{2}\pi(\hbar n)^3}{m_0(ke^2Z)^2}\right).
\end{align}
\begin{align}\label{1.17imk}
P=\left(\frac{\sqrt{2}hn^3}{m_0c^2\alpha^2}\right),
\end{align}
where $\alpha$ is the fine structure constant. Rydberg constant is given by,
\begin{align}\label{1.17iml}
R_\infty=\left(\frac{m_0c^2\alpha^2}{2hc}\right).
\end{align}
Substitution in Eq.~\eqref{1.17imk} gives orbital period of electron in hydrogen atom,
\begin{align}\label{1.17imm}
P=\left(\frac{n^3}{\sqrt{2}cR_\infty}\right),
\end{align}
where $n$ is the principle quantum number. Here periodic nature of time becomes evident. Time makes a quantum jump proportional to cube of the principle quantum number. Now $n$ is not just an integer constant but a function of orbital period of electron and therefore time dependent. And space dependent through relation $r=a_0n^2$.

In PR we are justified in proposing Eq.~\eqref{1.17imm} because unlike Schrodinger and Dirac theories, the orbital angular momentum in  PR is non-zero for $1s$ orbital. This is because we have succeeded in introducing the spin operator $\textbf{S}$ in the classical Laplacian \cite{77} from which we can obtain the expression
\begin{align}\label{2.36b}
(\boldsymbol{L}+\boldsymbol{S})^2\psi^{(2)}_{j,m_j}=\Gamma\hbar^2\psi^{(2)}_{j,m_j}.
\end{align}
\begin{align}\label{2.36bb}
\boldsymbol{J}^2\psi^{(2)}_{j,m_j}=j(j+1)\hbar^2\psi^{(2)}_{j,m_j}.
\end{align}
This allows us to compute the non-zero momentum $\sqrt{3/4}\hbar=0.866\hbar$ for $1s$ orbital. Thus we are justified in balancing the electromagnetic force with this circular motion for $1s$ orbital.
   
\section{Field equations in PR in presence of matter}

We have from Eqs.~\eqref{4.1a} and~\eqref{4.15},
\begin{align}\label{5.1}
(E-m_0c^2)=\Phi m_0=-\int \frac{\mu m}{r^2} \left(\cos{\psi}+\sin{\psi}\right)dr.
\end{align}
For cosmological application we are only interested in radial motions hence we take $\psi=0$. Secondly for small radial motions we assume $\gamma \approx const.$ which gives
\begin{align}\label{5.2}
mc^2-m_0c^2=\frac{\mu}{r} \gamma m_0,
\end{align}
\begin{align}\label{5.3}
\{1-(1/\gamma)\}c^2=\frac{\mu}{r},
\end{align}
The energy-momentum invariant Eq.~\eqref{1.4} gives
\begin{align}\label{5.4}
\gamma=(m/m_0)=\pm\{1-(v^2/c^2)\}^{-1/2},
\end{align}
Here the $\pm$ sign is due to the positive and negative energies of Dirac's theory. Introduction of Eq.~\eqref{5.4} in Eq.~\eqref{5.3} gives
\begin{align}\label{5.5}
c^2(1 \mp\{1-(v^2/c^2)\}^{1/2})=\frac{\mu}{r},
\end{align}
\begin{align}\label{5.7}
c^2-v^2=\left[\frac{\mu}{rc}-c\right]^2,
\end{align}
\begin{align}\label{5.8}
c^2dt^2-(dx^2+dy^2+dz^2)=\left[\left(\frac{\mu}{rc}\right)^2+c^2-\frac{2\mu}{r}\right]dt^2=ds^2,
\end{align}
Eq.~\eqref{5.8} is simply the flat Minkowski metric given by Eq.~\eqref{1.18} when $n=1$, and this equation is based on the conservation of energy equation~\eqref{5.1}. 
For application in cosmology we can introduce deviation factor $n$ in Eq.~\eqref{5.8} and then assuming $(\mu/rc)^2$ to be negligibly small, the general line element satisfying the Weyl postulate and the cosmological principle can be given by 
\begin{align}\label{5.6}
\begin{split}
ds^2&=c^2dt^2-na^2\left(\frac{dr^2}{1-Kr^2}+r^2d\theta^2
+r^2\sin^2{\theta}d\phi^2\right)
=\left[c^2-\frac{2\mu}{ar}\right]dt^2,
\end{split}
\end{align}
where $a(t)$ is the scale factor and parameter $K$ is equal to +1 or 0 or -1 as in Friedmann model and decides the curvature of 3-surfaces. All the observable evidence indicate that the universe is near flat corresponding to $K=0$, so we introduce this value in Eq.~\eqref{5.6} at the outset. This and the fact that each galaxy has a constant set of coordinates $(r,\theta,\phi)$, will considerably simplify the mathematics required for analyzing the model. For small and constant values of $n$, line element Eq.~\eqref{5.6} does satisfy Einstein's field equation $R_{\mu\nu}=0$. We can write this equation as 
\begin{align}\label{5.6a}
\begin{split}
\frac{2\mu}{ar}dt^2-na^2\left(dr^2+r^2d\theta^2
+r^2\sin^2{\theta}d\phi^2\right)=0.
\end{split}
\end{align}
This can be transformed to metric form as
\begin{align}\label{5.9}
g_{\mu\nu} dx^\mu dx^\nu=0, \qquad \text{where}
\end{align}
 \begin{align}\label{2.9g}
g_{\mu\nu}=
\begin{pmatrix}
	2\mu/ar & 0 & 0 & 0 \\
	0 & -a^2n & 0 & 0\\
	0 & 0 & -a^2nr^2 & 0\\
	0 & 0 & 0 & -a^2n(r^2\sin^2{\theta})
\end{pmatrix}.
\end{align}
where $dx^0=dt$, $dx^1=dr$, $dx^2=d\theta$, $dx^3=d\phi$.
The metric Eq.~\eqref{5.9} yields
\begin{align}\label{5.9l}
\nabla^2[g_{\mu\nu} dx^\mu dx^\nu]=0,
\end{align}
\begin{align}\label{5.9m}
\text{where} \quad \nabla^2=\frac{1}{r^2}\frac{\partial}{\partial r}\left(r^2\frac{\partial}{\partial r}\right)
\end{align}
In order to analyze the expanding universe scenario, we can use Eq.~\eqref{5.6a} for a small radial motion of the galaxy keeping $\theta$ and $\phi$ constant. This gives
\begin{align}\label{5.10}
\begin{split}
\frac{\mu}{ar}=\frac{na^2}{2}\left(\frac{dr}{dt}\right)^2=\frac{na^2v^2}{2}\equiv\frac{n(Hr)^2}{2}.
\end{split}
\end{align}
where $H=\dot{a}/a$ is Hubble parameter and deviation factor $n$ associated with this system can conform to GR provided we select
\begin{align}\label{5.11}
n=-\frac{1}{2}\left(1-\frac{\Lambda}{3H^2}\right).
\end{align}
Here dark energy \cite{38,39,74,75,78,80} associated with the cosmological constant $\Lambda$ is presumed to cause deviation in the flat Minkowski metric. In GR $\Lambda$ gets introduced through the action principle. Here $n$ is a unitless number. By introducing this deviation factor we are proposing that the presence of uniformly distributed dark energy on a cosmological scale can cause redshift of all the constituent particles of a galaxy. This is because the dark energy causes accelerated expansion of the universe which is bound to affect the redshift of every galaxy. This factor is not accounted by the weak field approximation and the corresponding deviation to the flat Minkowski metric in GR. Since PR relates the proper time of a body with the frequency shift of all the constituent particles of a body, we are justified in proposing the deviation factor Eq.~\eqref{5.11} which only alters the proper time interval of a galaxy without introducing any curvature. This deviation factor $n$ remains constant for any given epoch but varies from epoch to epoch because it is a function of the Hubble parameter. Therefore $n$ satisfies Einstein's field equation $R_{\mu\nu}=0$. 

For the field point within the source of gravitation, in accordance with Poisson's equation we get from Eq.~\eqref{5.10} and~\eqref{5.9m},
\begin{align}\label{5.12}
H^2-\frac{\Lambda}{3}=\frac{8}{3}\pi G\rho.
\end{align}
This is same as the Friedmann equation for flat universe \cite{74,75,78}. Hence the critical density in this model when $\Lambda=0$ comes out to be same as the Friedmann model
\begin{align}\label{5.13}
\rho_c=\frac{3H^2}{8\pi G}.
\end{align}
If we substitute $t=1/H$ and
\begin{align}\label{5.14}
c^2 \rho=(1/2)g \sigma T^4,
\end{align}
we get the time temperature relation
\begin{align}\label{5.15}
t=\left(\frac{3c^2}{16\pi Gg\sigma}\right)^{1/2}T^{-2},
\end{align}
where $t$ is the time of the epoch, $g$ the $g$ factor, $\sigma$ radiation constant, T the temperature.

If we take time derivative of Eq.~\eqref{5.10}, we get for constant $\Lambda$, the acceleration equation
\begin{align}\label{5.16}
\left(\frac{\ddot{a}}{a}\right)-\frac{\Lambda}{3}=\frac{2GM_0}{r^3},
\end{align}
For a point particle on a homogeneous sphere of radius $r$ and energy density $\rho$ Eq.~\eqref{5.16} reduces to
\begin{align}\label{5.17}
\left(\frac{\ddot{a}}{a}\right)-\frac{\Lambda}{3}=\frac{8}{3}\pi G\rho.
\end{align}
Positive sign on the right imply accelerated expansion.
Here we can introduce the equation of state $w=p/\rho$ through the relation
\begin{align}\label{5.18}
\rho \propto a^{-3(1+w)}.
\end{align}
which yields the relations
\begin{align}\label{5.19}
\dot{\rho}=-3H(\rho+p) \quad and \quad \dot{H}=-4\pi G(\rho+p).
\end{align}
If we compare Eqs.~\eqref{5.17} and~\eqref{5.12}, we find that $\dot{H}=0$, which means that Eq.~\eqref{5.17} is valid for $w=-1$. Therefore for other values of $w$, we can introduce Eq.~\eqref{5.19} in Eq.~\eqref{5.17} which gives
\begin{align}\label{5.20}
\left(\frac{\ddot{a}}{a}\right)-\frac{\Lambda}{3}=H^2+\dot{H}-\frac{\Lambda}{3}=-\frac{4}{3}\pi G(\rho+3p).
\end{align}
Therefore accelerated expansion occurs for $(\rho+3p)<0$. Since $H$ is constant for $w=-1$, we get the inflationary exponential expansion. 
\begin{align}\label{5.21}
a\propto e^{Ht}.
\end{align}
Looking at the above results we find that the theory is in conformance with the GR cosmology and the $\boldsymbol{\Lambda}$CDM model. For obtaining proper time interval of a galaxy we substitute Eq.~\eqref{5.11} for constant $n$ in Eq.~\eqref{1.1b} for the proper time interval where $v$ is to be replaced by $av=Hr$. This gives
\begin{align}\label{5.22}
d\tau=dt\left(1+\frac{r^2}{4c^2}\left(H^2-\frac{\Lambda}{3}\right)\right),
\end{align}
\begin{align}\label{5.23}
d\tau=dt\left(1+\frac{2 \pi G\rho r^2}{3c^2}\right).
\end{align}
Eq.~\eqref{5.23} is valid for small values of $v$ and this is where PR will differ from GR.

\section{Quantum gravity}
I have discussed the quantum gravity theory (PQGC) \cite{154} based on the periodic nature of time which has striking similarity to Friedmann equations. In this theory Hubble parameter is equated to Planck frequency at Planck epoch. Both have the units of frequency. So the universe begins with the first oscillations when Hubble parameter begins to roll down from a very high energy value which is Planck energy. This generates field of particles with Planck energy. There is no separate equation for scale factor and exponential expansion, but the exponential term of the wave equation itself induces the exponential expansion which is continuous till the present epoch. The term equivalent to the scale factor is provided by the familiar separated time dependent function $f(t)$ of the wave equation. The superscript of the exponential term can be written in terms of energy $E$, or Hubble parameter $H$ for curvature $K=0$. PQGC uses a modified Laplacian which includes particle spin parameter \cite{77}. The theory generates all the particle fields of the standard model from a single formula. The theory is based on unification of all particle fields into a single field at Planck epoch. Scale factor $a$ and cosmological constant $\Lambda$ are included in separated time dependent function $f(t)$ which can be seen through following relation. In PQGC \cite{154} we have derived expression,
\begin{align}\label{c7c}
\frac{\ddot{f}}{f}=-\frac{B1}{\hbar \kappa^2 C1} \frac{8\pi G \boldsymbol{\rho}}{3}.  
\end{align} 
If we substitute Eq.~\eqref{5.17} in Eq.~\eqref{c7c} we get,
\begin{align}\label{c7ca}
\frac{\ddot{f}}{f}=\frac{B1}{\hbar \kappa^2 C1}\left[\frac{\Lambda}{3}-\left(\frac{\ddot{a}}{a}\right)\right].  
\end{align}
From Eq.~\eqref{c7ca} we can see that the expansion of universe is due to quantum field expanding and not due to the space expanding as in the GR. Dark energy (68.3$\%$) is the energy of quantum field which is not manifest as the  particles, or manifest as hard to detect ground state particles. Quantum fields are expanding within the singular motionless fundamental substance of the universe which is infinite in extent, but the quantum fields are not infinite. So the zero point energy of the quantum fields can be restricted to match the cosmological constant as in Eq.~\eqref{c7ca} and this should solve the cosmological constant problem. The presence of the motionless infinite fundamental substance of the universe sets the same initial conditions everywhere including causally disconnected regions of space which explain the cosmological Horizon Problem. The theory works with or without the dark energy. James Web Space Telescope (JWST) has discovered large density of galaxies at high redshifts which puts tight constraints on the expansion history of the universe and rule out the major portion of the parameter space of the dark energy models \cite{162}.

\subsection{Gravitation within electromagnetic wave} \label{gemw}

General relativity derivation of gravitational frequency shift of light does not provide the quantum mechanism of the frequency shift. However with periodic time in PR it is possible to explain this quantum mechanism as follows. 
In an electromagnetic wave there is a stream of successive wavelets traveling at constant velocity of light. Each wavelet having energy equal to a quanta called photon. The mass equivalent of this energy is given by $E/c^2$. If we consider the gravitational attraction between two successive photons within the electromagnetic wave then we can write the expression for gravitational force between them as
\begin{align}\label{1.22a}
\mathbf{F}=\frac{G}{\lambda^2}\left(\frac{E}{c^2}\right)^2\mathbf{\bar{v}} =\left(m\mathbf{a}+\frac{h\mathbf{v}}{c^2}\frac{d\nu}{dt}\right),	
\end{align}
where $G$ is the gravitational constant and $\lambda$ the distance between two successive photons.
For light the classical acceleration $\mathbf{a}=0$ and $\mathbf{v}=\mathbf{c}$, but the de Broglie force is not zero. Eq.~\eqref{1.22a} reduces to
\begin{align}\label{1.22b}
F=\frac{G}{\lambda^2}\left(\frac{E}{c^2}\right)^2=\left(\frac{h}{c}\frac{d\nu}{dt}\right),	
\end{align}
Here the gravitational force $F$ on the left is always attractive but r.h.s. can be positive or negative depending on whether light is red shifted or blue shifted. This is like two binary stars ejecting lot of energy and moving away from each other or absorbing lot of energy and coming close to each other. In both cases gravity is always attractive.
Substituting $\lambda=cT=c/\nu$ gives,
\begin{align}\label{1.22c}
\frac{Gh}{c^5}=\frac{1}{\nu^4}\frac{d\nu}{dt}.	
\end{align}
Now we can safely replace linear time increment $dt$ with the periodic time increment $dT=1/d\nu$.
\begin{align}\label{1.22d}
\frac{Gh}{c^5}=\frac{1}{\nu^4}(d\nu)^2.	
\end{align}
Substituting $h=2\pi \hbar$ we get,
\begin{align}\label{1.22e}
\nu=\frac{1}{\sqrt{2\pi}}\left(\frac{d\nu}{\nu}\right)\sqrt{\frac{c^5}{G\hbar}},	
\end{align}
where quantity in the bracket is just the redshift factor $z$ as defined in the standard model of big bang \cite{74}. However, what we have here is the gravitational redshift and not the Doppler shift of Hubble's law. Hence
\begin{align}\label{1.22f}
2\pi\nu=\sqrt{2\pi}z\sqrt{\frac{c^5}{G\hbar}}.
\end{align}

In our earlier work \cite{154}, we defined quantity $\alpha$ as the inverse of the Planck length $l_p$ and the Hubble parameter $H$ in terms of the Planck angular frequency $\omega_p$
\begin{align}\label{b33}
\alpha=\sqrt{\frac{c^3}{G\hbar}}=\frac{1}{l_p}\ ,
\end{align}
\begin{align}\label{b34}
H=\kappa \omega_p=\kappa \alpha c=\kappa \sqrt{\frac{c^5}{G\hbar}}\ ,
\end{align}
where we defined $\kappa$ with an ansatz that it is a unitless real number. Hubble parameter has the units of frequency. Earlier \cite{154} we had defined Planck energy as,  
\begin{align}\label{b38}
E_p=\hbar \alpha c=\hbar \omega_p.
\end{align}
As a final solution to the wave equation, particle energy levels were given by 
\begin{align}\label{b66}
E=E_p\sqrt{\frac{B1}{C1}},
\end{align}
where $B1$ and $C1$ are unitless numbers. In case of electromagnetic wave, $E=h\nu$. Inserting this in Eq.~\eqref{b66} we get,
\begin{align}\label{b66a}
H=2\pi \kappa\nu\sqrt{\frac{C1}{B1}},
\end{align}

\begin{align}\label{b66b}
\frac{d\nu}{\nu}=\sqrt{\frac{B1}{2\pi C1}}=\sqrt{\frac{Gh}{c^5}}\nu.
\end{align}

From Eq.~\eqref{b66a} we see that the Hubble parameter is a function of particle frequency and from Eq.~\eqref{b66b} we see that the gravitational frequency shift is a function of unitless variables $B1$ and $C1$ and proportional to frequency of light at any given time. So here we have the robust relativistic derivation of the variable gravitational redshift as the Hubble parameter rolls down from Planck epoch to the present epoch. This is done using de Broglie force, periodic time and gravitational attraction between two successive photons within the electromagnetic wave which merge with each other at about $10^{-26}$ sec. before the big crunch in PQGC \cite{154} terminating the motion of the wave meaning that the wave will collapse. In PQGC it is not possible to generate photon parameters before this time. This is where gravity unites with the electromagnetic force. Same phenomenon occurs with Savitons at Planck energy when the Saviton waves collapse during big crunch. Gravitational constant $G$ in Eqs.~\eqref{b33} and~\eqref{b34} does not come from a quantum gravity theory, but it comes through Einstein's field equations. So what we have here is the first theoretical evidence of existence of gravitational forces within the electromagnetic wave. 

As we change the value of $\kappa$ by hand, it changes the Hubble parameter and we know from general relativity that the energy density of the universe is proportional to square of the Hubble parameter. Energy density of the universe is continuously changing due to the expansion of the universe. In the lower energy density epochs, particles also have lower energies and so does photons of electromagnetic waves. Gravitational redshift discussed in this formalism is intrinsic to electromagnetic waves. Lower the energy of the wave, lower is the intrinsic gravitational redshift. Therefore at present epoch this gravitational redshift is very small. This is due to very small value of the Hubble parameter at present epoch which is comparable to the small value of the cosmological constant.

From these calculations we can conclude that the electromagnetic wave is held together by gravitational forces. Not only that, even the seven monochromatic wave frequencies are held together by gravitational forces as a packet of white light. When these gravitational forces are overcome by processes like refraction, diffraction or polarization, then the electromagnetic wave decomposes into components like photon particles or seven colors of the rainbow etc. The monochromatic waves of light are held together due to attraction between relativistic masses $E/c^2$ of different frequencies. This is gravity assisted wave stacking. Here the stacked monochromatic waves are incoherent unlike the laser beam because they have different frequencies. The laser beam is also a case of gravity assisted stacked light waves having same phase and hence coherent.   

Mechanism of gravitational frequency shift of light can thus be explained as due to absorption or ejection of quanta of energy from or to the surroundings. When photons absorb gravitational energy, two successive photons come closer due to greater gravitational attraction as per Newton's inverse square law and thus light gets blue shifted. When photon loses energy to the surroundings, the gravitational attraction between two successive photons is reduced which stretches the wavelength and the light gets red shifted. All this happens at a constant velocity of light. 

Whenever GR physicists use Lorentz force as the rate of change of momentum, Eq.~\eqref{1.22}, they are ignoring the de Broglie force and the periodic time. This is because GR has abandoned the relativistic mass and thus further differentiation of momentum is not possible. This eliminates the de Broglie force and saves the principle of equivalence. As a consequence, both the principle of equivalence and the gravitational redshift are not invariant in GR. But in PR, relativistic mass is part of the theory and the gravitational mass is equal to relativistic mass. This makes both, the principle of equivalence as well as the gravitational redshift invariant.

One should be careful in using Eq.~\eqref{1.22c} because it is partly based the on quantum distance $\lambda$ and partly on linear time $t$. They don't go together. One may get tempted to write,
\begin{align}\label{1.22g}
\int dt=\frac{c^5}{Gh} \int \frac{1}{\nu^4}d\nu,	
\end{align} 
but it will not give you any meaningful results. It has to be either linear time and linear distance which we used earlier in Eq.~\eqref{4.4} for deriving gravitational redshift of light coming from the surface of the sun or the periodic time and periodic distance as we used above in Eq.~\eqref{1.22d}.

If we plug in gamma ray photon data in Eq.~\eqref{b66b} we get gravitational redshift of about $1.13\times10^{-22}$ at $10^{-26}$ sec. after big bang which is very small. If we plug in Saviton data which is a massless boson having Planck energy then we get gravitational redshift of $0.4$ at Planck time which is significant. This gravitational redshift occurs over a distance of single wavelength. This is not an astronomical distance. This is the kind of gravitational radiation coming out from about $10^{62}$ Savitons at Planck time which created the entire universe. So all the known fundamental particles of physics are products of this gravitational redshift. Published version of Section~\ref{gemw} is available in \cite{222}.    

\subsection{Quartic law of quantum gravity}

We can write Eq.~\eqref{1.22b} as
\begin{align}\label{1.22j}
F=\frac{G}{\lambda^2}\left(\frac{E}{c^2}\right)^2.	
\end{align}
Substituting $\lambda=cT=c/\nu$ gives,
\begin{align}\label{1.22k}
F=\frac{Gh^2\nu^4}{c^6}=\frac{GE^4}{h^2c^6}.	
\end{align}
Eq.~\eqref{1.22k} shows that the gravitational force is a quartic function of energy or frequency and $F\propto E^4$ or $F\propto \nu^4$.\\
We can write Eq.~\eqref{1.22f} as
\begin{align}\label{1.22h}
E=z\sqrt{\frac{c^5h}{G}}.
\end{align}
Here energy $E$ can be positive or negative depending on redshift or blueshift. Positive only means light wave is throwing out energy to the surrounding and negative means energy is absorbed from the surrounding.
Substitution of Eq.~\eqref{1.22h} in Eq.~\eqref{1.22k} gives,
\begin{align}\label{1.22ha}
F=z^4\sqrt{\frac{c^4}{G}}.
\end{align}
In Eq.~\eqref{1.22ha}, regardless of redshift or blueshift, the gravitational force $F$ is always positive and attractive. The formalism developed for the electromagnetic wave above can be extended to the first massless bosons created at Planck epoch, namely savitons in PQGC theory \cite{154}. After introduction of gravity in electromagnetic wave, the main difference between photons and savitons is the amount of energy they carry. Savitons have Planck energy and contains energy required to create strong force. The maximum amount of energy that savitons can carry is limited by its smallest possible wavelength which is the Planck length. This corresponds to a specific particle frequency limited by the relation $\nu_{max}=c/\lambda_{min}$. When saviton acquire more energy then allowable, the particle wave will collapse. Two successive savitons in a wave will unite with each other and frequency $\nu$ will drop to zero. Both the energy and gravity will disappear and saviton will become motionless and unite with the fundamental motionless substance of the universe which is the singularity. The process is reversible.

\subsubsection{How elementary particles acquire mass and charge}

In PQGC \cite{154}, mass is acquired by a particle when part of its kinetic energy gets transformed into potential energy so the velocity of the particle drops and inertia sets in. Higgs boson also acquire its mass in this way. The first particles created in this theory were bosons called Savitons having Planck energy and massless like photons and having velocity of light. In this, energy gets condensed into mass like water getting condensed into ice. This is phase transformation of energy into mass. So the particles acquire their mass through phase transformation of their own energy. Gravitational redshift described in Eq.~\eqref{1.22h} plays the principle role in creation of new particles. All the matter of the universe is created through this red shifting. In a reverse process of blue shifting, all the matter of the universe is absorbed by savitons in a massless state which then turns around to acquire Planck energy and ultimately unite with the fundamental motionless substance of the universe.

Charge is acquired as follows. We know that electron positron pair can annihilate into two photons. This process is reversible. When photons transform into electron positron pair, they acquire their mass as described above and the electric field associated with the electromagnetic wave, condenses into charge. So this is also a phase transformation of the energy of the electric field. The process is similar to Bose Einstein Condensate.

\subsubsection{Predominance of matter over anti-matter}

In PQGC, universe begins with the creation of the first particle ($10^{62}$ particles) we call Saviton which is a massless boson having Planck energy at Planck time. Like photon, saviton is its own anti-particle. So the universe was matter dominated right from the beginning. Savitons decayed into other particles through gravitational red shifting. A small fraction of these decayed particles developed matter anti-matter division with opposite spin and opposite charge. All quarks and anti-quarks except top quark remained confined in pairs due to strong force of gluons. Few like electron and positron remained free. It is possible that the concept of dipole gravity \cite{161} in conjuction with the relativistic mass along with the radial component of Lense-Thirring force can explain creation of particle anti-particle pairs having Planck energy at Planck time from a single particle of Planck length rotating at a very high speed? This can naturally cause inflation and the expansion of the universe. In this, rotational kinetic energy of dipole gravity particle is transformed into rectilinear kinetic energy of monopole gravity particles. 

\section{Gravitational Waves}\label{GW}
The orbital period derivative of a binary pulsar is calculated and verified with great accuracy \cite{20}. The most popular explanation for the resulting energy loss of the binary system is given in terms of the emission of gravitational waves \cite{62,63,64,65}. Recently LIGO scientists detected the gravitational waves arising from binary black hole merger \cite{129}. But no gravitational waves arising from the binary pulsars have yet been detected \cite{127}. Under the circumstances, it would be useful to look at some what modified explanation for the cause of the orbital period decay of binary pulsars. PR relates the orbital period of the pulsar with the period of the phase of the constituent particles of the pulsar. The idea is that, when small amount of gravitational energy is released, it may not generate gravitational waves, but this energy can alter the period of the phase of the constituent particles of the pulsar. Constituent particles of the pulsar are in bound states in atoms and molecules. They have their orbital motions around their nucleus. At the same time, each particle also has a somewhat rectilinear motion along the direction of the orbital velocity of the pulsar. This later motion has the de Broglie phase wave associated with it. Decay of the orbital period of the pulsar causes decay of the period of this phase of the constituent particles of the pulsar. So the gravitational energy lost by the pulsar is equal to the energy gained by the constituent particles of the pulsar. This results in increased kinetic energy of the pulsar. So the potential (gravitational) energy is converted into kinetic energy. When very large amount of gravitational energy is released, then only it will generate gravitational waves, like in case of binary black holes. This is because the large amount of energy released destroys the bonds between the constituent particles of the orbiting body first, and then the excess energy finds its way as the gravitational waves.

\subsection{Gravitational redshift of gravitational Waves}
Xian et. al. \cite{157} have discussed mass-redshift degeneracy with respect to cosmological, Doppler and gravitational redshift of gravitational waves arising from merger of binary black holes (BBHs) in the vicinity of a supermassive black hole (SMBH). Here the authors have used the standard gravitational redshift formula of GR which is not invariant.
\begin{align}\label{4.8}
1+Z_g=\left(1-\frac{R_s}{r}\right)^{-\frac{1}{2}}.
\end{align}
where $R_s$ is the Schwarzschild radius and $r=\Delta$ is the radius of the body. Eq.~\eqref{4.8} reduces to 
\begin{align}\label{4.9}
Z_g^{GR}=\left(\frac{GM}{c^2\Delta}\right).
\end{align}
Earlier we have derived gravitational redshift formula in Section~\ref{GR} which is invariant and given by
\begin{align}\label{4.5a}
\frac{\mu}{c^2}\frac{l}{(\Delta^2+\Delta l)}=\frac{\varphi_1-\varphi_2}{c^2}=\ln\left(1-\frac{\delta\nu}{\nu_s}\right),
\end{align} 
For very large value of $l$ Eq.~\eqref{4.5} reduces to
\begin{align}\label{4.10a}
Z_g^{PR}=exp\left(\frac{GM}{c^2\Delta}\right)-1,
\end{align}
and for a very small value of the quantity in the bracket like in case of gravitational redshift of light coming from the sun, Eq.~\eqref{4.10a} reduces to Eq.~\eqref{4.9}. But in case of gravitational redshift of gravitational waves arriving from BBH merger, Eq.~\eqref{4.10a} cannot be ignored because it gives greater redshift in strong field than the GR formula which is based on weak field approximation. In case of SMBHs, the difference will be even more striking. If we use the GW150914 data \cite{129}; $m1=36M_\odot$, $m2=29M_\odot$, we get the chirp mass $\mathcal{M}=28.09M_\odot$. We use the Schwarzschild radius for the sum of the component masses $M=(m_1+m_2)$ to be $R_s=\Delta=192.01$ Km. This gives $Z_g^{GR}=0.2161$ and $Z_g^{PR}=0.24125$. If we compute the emitted frequency $f_e$ at source based on the observed frequency at peak amplitude $f_o=150$ Hz, we get $f_e^{GR}=191.35$ Hz and $f_e^{PR}=197.69$ Hz. Hence prediction of PR is $6.34$ Hz higher than GR in the strong field regime of GW150914. This difference can become even more striking in case of SMBHs. In PR, the event horizon is at $1R_g$, but we have used Schwarzschild radius here for comparison purpose. In PR, $R_g$ is defined with total mass which is relativistic mass including spin energy. If irreducible mass is used for defining $R_g$ then the event horizon is between $1R_g$ and $\sqrt{2}R_g$. The event horizon of Kerr black hole is between $1R_g$ and $2R_g$.    

\subsection{Relativistic and rest masses of graviton in periodic relativity}
For a particle having only relativistic mass $m$ (rest mass $m_0=0$), traveling radially in the gravitational field of mass M at velocity of light, we defined the invariant relation between Newtonian force and the de Broglie force in Section~\ref{GR} as given below. Using this equality we derived the formula for gravitational redshift of light given by Eq.~\eqref{4.5a}. 
\begin{align}\label{4.11a}
\frac{\mu m}{r^2}=\frac{h}{c}\frac{d\nu}{dt},
\end{align}
where $\mu=GM$, $m$ is the relativistic mass traveling at speed of light in the gravitational field of mass $M$, $h$ is Planck constant, $c$ is the velocity of light. $\nu$ is the de Broglie frequency of the particle wave. Using this equality of Eq.~\eqref{4.11a} we will define the deviation factor $n$ as a ratio of Newtonian acceleration to de Broglie acceleration for the PR line element given by Eq.~\eqref{1.17} which can now become a wave equation because of the presence of the frequency term and Compton wavelength. For null geodesic we have $v=c$ and $n=1$.  
\begin{align}\label{4.12a}
n=1=\frac{GM}{r^2}\left(\frac{h}{mc}\frac{d\nu}{dt}\right)^{-1},
\end{align}
We can relate Eq.~\eqref{4.12a} with GW150914 by substituting $M=(m_1+m_2)$, $r=1R_g=GM/c^2$, and $\dot{\nu}=\dot{f}$.
Then for mass of graviton in the source frame we get
\begin{align}\label{4.13a}
m=m_{gr}=\frac{R_g^2h\dot{f}}{cG(m_1+m_2)}=\frac{hG(m_1+m_2)\dot{f}}{c^5},
\end{align} 
where $\dot{f}$ is given by 
\begin{align}\label{4.14a}
\frac{df}{dt}=\frac{96}{5}\pi^{\frac{8}{3}}\left(\frac{G\mathcal{M}}{c^3}\right)^{\frac{5}{3}}f^{\frac{11}{3}},
\end{align} 
where chirp mass $\mathcal{M}=28.09M_\odot$ and $f=35$ Hz. This gives relativistic mass of graviton $m_{gr}=1.02\times 10^{-33}$ eV/$c^2$. If the frequency of the gravitational waves is such that $hf>>m_gc^2$, where $m_g$ is the rest mass of graviton, then the bound on the graviton Compton wavelength $\lambda_g=h/m_gc$ \cite{171}, can be given by 
\begin{align}\label{4.14ab}
\lambda_g > 3\times 10^{12}\text{Km}\left(\frac{D}{200\,\text{Mpc}}\frac{100\,\text{Hz}}{f}\right)^{1/2}\left(\frac{1}{f\Delta t}\right)^{1/2}.
\end{align} 
From Eq.~\eqref{4.13a}, we get $f=m_{gr}c^2/h=2.21\times 10^{-2}$\,Hz. We have $D=410$\,Mpc and $f\Delta t=\rho^{-1}=1/10$. Substituting these values in Eq.~\eqref{4.14ab}, we get $\lambda_g > 9.13\times 10^{14}$\,Km. This gives bound on the graviton rest mass $m_g=h/(\lambda_g c)<1.51\times 10^{-41}$eV/$c^2$, which is well within the currently accepted limit $m_g<6.76\times 10^{-23}$eV/$c^2$ \cite{178}.

\subsection{Gravitational wave equation in periodic relativity}
From Eq.~\eqref{4.3a} we can write for $n=1$,
\begin{align}\label{4.15a}
\frac{d(h\nu)}{dt}=\frac{GM}{r^2}(mc).
\end{align}
\begin{align}\label{4.16a}
\frac{dE}{dt}=\frac{GM}{r^2}(\textbf{p}^{2})^{1/2}.
\end{align}
We replace $E$ by the differential operator $i\hbar\partial/\partial t$, and $\textbf{p}$ by $-i\hbar\nabla_j$ that act on the wave function $\psi$. The spin dependent $\nabla_j$ operator is discussed at length in \cite{77}.
\begin{align}\label{4.17a}
\frac{d}{dt}\left(i\hbar\frac{\partial \psi}{\partial t}\right) =\frac{GM}{r^2} [(-i\hbar\nabla_j)^2]^{1/2} \psi.
\end{align}
Eq.~\eqref{4.17a} is valid for orbital and radial motion of both gravitational waves as well as electromagnetic waves. Assuming spherically symmetric potential, we define operator $\boldsymbol{\nabla}_j^2$ in spherical polar coordinates as
\begin{align}\label{2.21p}
\boldsymbol{\nabla}_j^2=\frac{1}{r^2}\left[\frac{\partial}{\partial r}\left(r^2\frac{\partial}{\partial r}\right)-\frac{1}{\hbar^2}(\boldsymbol{L}+\boldsymbol{S})^2\right], 
\end{align}
where $\boldsymbol{L}$ is the orbital angular momentum operator, and
$\boldsymbol{S}$ is the spin angular momentum operator. The wave equation reduces to
\begin{align}\label{4.18a}
\frac{d}{dt}\left(\frac{\partial \psi}{\partial t}\right) =\frac{GM}{r^2} (\nabla_j^2)^{1/2} \psi.
\end{align}
Note that the use of ratio of accelerations eliminates the need for potential energy term commonly used in quantum mechanics. We consider a particular solution of Eq.~\eqref{4.18a} that can be written as a product $\psi(\boldsymbol{r},t) = z(\boldsymbol{r}) f(t)$. A general solution can be written as a sum of such separated solutions. If we substitute the above product in Eq.~\eqref{4.18a} and divide through by the product, we get 
\begin{align}\label{10.19}
\frac{d}{dt}\left(\frac{1}{f}\frac{df}{dt}\right) =\frac{1}{z}\frac{GM}{r^2} (\nabla_j^2)^{1/2} z.
\end{align}
If we define another function $u(\mathbf{r})$ such that
\begin{align}\label{10.20}
\frac{1}{z}\frac{GM}{r^2} (\nabla_j^2)^{1/2} z=\frac{GM}{r^2} \left(\frac{1}{u}\nabla_j^2 u \right)^{1/2},
\end{align}
then we can write Eq.~\eqref{10.19} as
\begin{align}\label{10.21}
\frac{d}{dt}\left(i\hbar\frac{1}{f}\frac{df}{dt}\right) =i\hbar\frac{GM}{r^2} \left(\frac{1}{u}\nabla_j^2 u \right)^{1/2}.
\end{align}
Since the left side depends only on $t$ and right side only on $\boldsymbol{r}$, both side must be equal to a same separation constant which in this case is $dE/dt$. Then equation for $f$ can be easily integrated to give,
\begin{align}\label{10.22}
f(t)=C\:\exp(-iEt/\hbar), 
\end{align}
where $C$ is an arbitrary constant and the equation for $u$ becomes,
\begin{align}\label{10.23}
\frac{dE}{dt}=i\hbar\frac{GM}{r^2} \left(\frac{1}{u}\nabla_j^2 u \right)^{1/2},
\end{align}
\begin{align}\label{10.24}
\dot{\omega}^2 u(\boldsymbol{r})=-\left(\frac{GM}{r^2}\right)^2 \nabla_j^2 u(\boldsymbol{r}).
\end{align}
where $E=h\nu$ and $\dot{\omega}=2\pi\dot{\nu}$. Substitution of Eq.~\eqref{2.21p} in Eq.~\eqref{10.24} gives,
\begin{align}\label{10.25}
 \begin{split}
&\left[\left(\frac{\dot{\omega}r^3}{GM}\right)^2+\frac{\partial}{\partial r}\left(r^2\frac{\partial}{\partial r}\right)\right]u(\boldsymbol{r})\\&=\left[i\left(\frac{1}{\sin\theta}\frac{\partial}{\partial \theta}\left(\sin\theta\frac{\partial}{\partial\theta}\right)+\frac{1}{\sin^2\theta}\frac{\partial^2}{\partial\phi^2}\right)^\frac{1}{2}+\frac{\boldsymbol{S}}{\hbar}\right]^2 u(\boldsymbol{r}).
 \end{split}
\end{align}
The radial and the angular parts can then be separated by substituting 
\begin{align}\label{10.26}
u(r,\theta,\phi)=R(r)Y(\theta,\phi)
\end{align}
in Eq.~\eqref{10.25} and dividing through by RY.
\begin{align}\label{10.27}
 \begin{split}
&\left(\frac{\dot{\omega}r^3}{GM}\right)^2+\frac{1}{R}\frac{\partial}{\partial r}\left(r^2\frac{\partial R}{\partial r}\right)\\&=\frac{1}{Y}\left[i\left(\frac{1}{\sin\theta}\frac{\partial}{\partial \theta}\left(\sin\theta\frac{\partial Y}{\partial\theta}\right)+\frac{1}{\sin^2\theta}\frac{\partial^2 Y}{\partial\phi^2}\right)^\frac{1}{2}+\frac{\boldsymbol{S}}{\hbar}\sqrt{Y}\right]^2.
 \end{split}
\end{align}
Since the left side of Eq.~\eqref{10.27} depends only on $r$ and the right side depends only on $\theta$ and $\phi$, both sides must be equal to a constant that we call $\Gamma$. Thus Eq.~\eqref{10.27} gives us a radial equation 
\begin{align}\label{10.28}
\frac{\partial}{\partial r}\left(r^2\frac{\partial R}{\partial r}\right)+\left(\frac{\dot{\omega}r^3}{GM}\right)^2R-\Gamma R=0,
\end{align}
and an angular equation
\begin{align}\label{10.29}
i\left(\frac{1}{\sin\theta}\frac{\partial}{\partial \theta}\left(\sin\theta\frac{\partial Y}{\partial\theta}\right)+\frac{1}{\sin^2\theta}\frac{\partial^2 Y}{\partial\phi^2}\right)^\frac{1}{2}=\left(\sqrt{\Gamma Y}-\frac{\boldsymbol{S}}{\hbar}\sqrt{Y}\right)=\sqrt{\lambda Y}.
\end{align}
From \cite{77} we know that $\Gamma=j(j+1)$, $\boldsymbol{S}/\hbar=\sqrt{s(s+1)}$ and $\sqrt{\lambda}=\sqrt{l(l+1)}$. Hence,
\begin{align}\label{10.30}
\frac{1}{\sin\theta}\frac{\partial}{\partial \theta}\left(\sin\theta\frac{\partial Y}{\partial\theta}\right)+\lambda Y =-\frac{1}{\sin^2\theta}\frac{\partial^2 Y}{\partial\phi^2}.
\end{align}
The angular equation can be further separated by substituting 
\begin{align}\label{10.31}
Y(\theta,\phi)=\Theta(\theta)\Phi(\phi)
\end{align}
and dividing by $\Theta\Phi$.
\begin{align}\label{10.32}
\biggl[\frac{1}{\Theta}\frac{1}{\sin\theta}\frac{d}{d\theta}\left(\sin\theta\frac{d\Theta}{d\theta}\right)+\lambda\biggr]\sin^2\theta
=-\frac{1}{\Phi}\frac{d^2\Phi}{d\phi^2}=\eta.
\end{align}
Hence we end up with two equations of Schr$\ddot{o}$dinger theory.
\begin{align}\label{10.33}
\frac{d^2\Phi}{d\phi^2}+\eta\Phi=0,
\end{align}
\begin{align}\label{10.34}
\frac{1}{\sin\theta}\frac{d}{d\theta}\left(\sin\theta\frac{d\Theta}{d\theta}\right)+\left(\lambda-\frac{\eta}{\sin^2\theta}\right)\Theta=0.
\end{align}
Eq.~\eqref{10.33} have the same solution as that given by the Schr$\ddot{o}$dinger theory where $\eta$ is chosen to be equal to square of an integer $m$ which takes on positive or negative integer values or zero. Therefore,
\begin{align}\label{10.35}
\Phi_m(\phi)=(2\pi)^{-\frac{1}{2}}\:\exp(im\phi).
\end{align}
We confine the motion of gravitational waves (or electromagnetic waves) in $z-x$ plane by putting $\phi=0$. Substituting $\phi=0$ in Eq.~\eqref{10.35} gives,
\begin{align}\label{10.36}
\Phi=\frac{1}{\sqrt{2\pi}}.
\end{align}
Therefore Eq.~\eqref{10.33} can be satisfied only if we put $\eta=0$. This reduces Eq.~\eqref{10.34} to 
\begin{align}\label{10.37}
\frac{1}{\sin\theta}\frac{d}{d\theta}\left(\sin\theta\frac{d\Theta}{d\theta}\right)+\lambda\Theta=0.
\end{align}
Differentiating the first term brings it to the form,
\begin{align}\label{10.38}
\tan\theta\frac{d^2\Theta}{d\theta^2}+\frac{d\Theta}{d\theta}+\lambda\tan\theta\Theta=0.
\end{align}

\subsubsection{Gravitational-wave strain amplitude using quantum mechanical formalism}
Eq.~\eqref{10.38} is a second order homogeneous linear equation. We seek a particular solution of Eq.~\eqref{10.38} by introducing the initial condition $\theta=0$. This gives
\begin{align}\label{10.39}
\frac{d\Theta}{d\theta}=0.
\end{align} 
Therefore we get the first particular solution
\begin{align}\label{10.40}
\Theta_1=C_3=Constant.
\end{align} 
If we can find another particular solution $\Theta_2$ of Eq.~\eqref{10.38} so that both solutions are linearly independent, then the general solution can be expressed as
\begin{align}\label{10.41}
\Theta=C_1\Theta_1+C_2\Theta_2,
\end{align} 
where $C_1$ and $C_2$ are arbitrary constants. The coefficients of three terms on the left of Eq.~\eqref{10.38} can be designated $a_0$, $a_1$ and $a_2$. In order to use Liouville's formula for finding the second particular solution, it is required that $a_0=1$ and also, the solution can be very simplified if we have $a_1=1$, which we already have. The first condition can be met only for those values of $\theta$ for which $\tan\theta=1$. So we proceed with the derivation with a remark that our general solution of Eq.~\eqref{10.38} will be valid only for $\theta=45^\circ$ and $\theta=225^\circ$. However, it should be possible to derive general solution applicable to all values of $\theta$.

By virtue of Liouville's formula, we can write
\begin{align}\label{10.42}
\Theta_2'\Theta_1-\Theta_2\Theta_1'=Cexp(-\int a_1 d\theta).
\end{align}  
Dividing both sides by $\Theta_1^2$ we get
\begin{align}\label{10.43}
\frac{\Theta_2'\Theta_1-\Theta_2\Theta_1'}{\Theta_1^2}=\frac{1}{\Theta_1^2}Cexp(-\int a_1 d\theta).
\end{align} 
For $a_1=1$ and $\Theta_1=C_3$ we get
\begin{align}\label{10.44}
\frac{d}{d\theta}\left(\frac{\Theta_2}{\Theta_1}\right)=\frac{1}{C_3^2}Cexp(-\theta),
\end{align} 
\begin{align}\label{10.45}
\frac{\Theta_2}{\Theta_1}=\int\frac{1}{C_3^2}Cexp(-\theta)d\theta+C'.
\end{align} 
For getting a particular solution, we can put $C'=0$ and $C=1$.
\begin{align}\label{10.46}
\Theta_2=\int\frac{1}{C_3}exp(-\theta)d\theta,
\end{align} 
\begin{align}\label{10.47}
\Theta_2=\frac{-1}{C_3}exp(-\theta).
\end{align} 
Comparing Eq.~\eqref{10.47} with Eq.~\eqref{10.40} we can see that $\Theta_1$ and $\Theta_2$ are linearly independent solutions. Hence, the general solution of Eq.~\eqref{10.38} is given by
\begin{align}\label{10.48}
\Theta=C_1C_3-\frac{C_2}{C_3}exp(-\theta).
\end{align} 
By putting $C_4=C_1C_3$ and $C_5=C_2/C_3$ we can write
\begin{align}\label{10.49}
\Theta=C_4-C_5exp(-\theta).
\end{align} 
Substitution of Eq.~\eqref{10.49} in Eq.~\eqref{10.38} gives
\begin{align}\label{10.50}
\left(\frac{C_4}{C_5}\right)=exp(-\theta)\left(\frac{1}{\lambda}+1-\frac{1}{\lambda\tan\theta}\right).
\end{align}
Remembering that our solution is valid only for $\tan\theta=1$, we get
\begin{align}\label{10.51}
\left(\frac{C_4}{C_5}\right)=exp(-45^\circ).
\end{align}
The normalization condition for the angular wave function $\Theta$ in Eq.~\eqref{10.49} is given by
\begin{align}\label{10.52}
\int|\Theta|^2d\theta=\int(C_4-C_5exp(-\theta))^2d\theta=1.
\end{align}  
Integration of Eq.~\eqref{10.52} gives
\begin{align}\label{10.53}
C_4^2\theta-2C_5^2exp(-2\theta)+C'+2C_4C_5exp(-\theta)+C''=1,
\end{align}  
where we put constants of integration $C'=0$, $C''=0$ and divide through by $C_5^2$.
\begin{align}\label{10.54}
\left(\frac{C_4}{C_5}\right)^2\theta-2exp(-2\theta)+2\left(\frac{C_4}{C_5}\right)exp(-\theta)=\frac{1}{C_5^2}.
\end{align} 
Substituting Eq.~\eqref{10.51} in Eq.~\eqref{10.54} and introducing initial condition as $\theta=0$, we get
\begin{align}\label{10.55}
C_5=\frac{1}{\sqrt{2}}(exp(-45)-1)^{-1/2}.
\end{align}
We can write Eq.~\eqref{10.49} as
\begin{align}\label{10.56}
\Theta=C_5\left(\frac{C_4}{C_5}-exp(-\theta)\right).
\end{align} 
Substituting Eq.~\eqref{10.51} and Eq.~\eqref{10.55} in Eq.~\eqref{10.56} we get
\begin{align}\label{10.57}
\Theta=\frac{1}{\sqrt{2}}(exp(-45)-1)^{-1/2}\left(exp(-45)-exp(-\theta)\right).
\end{align}
\begin{align}\label{10.58}
\Theta=\frac{1}{i\sqrt{2}}\left[(exp(-45)-exp(-\theta)+\frac{1}{2}exp(-45)^2-\frac{1}{2}exp(-45)exp(-\theta)\right].
\end{align}
Last two terms in the bracket are very small in around $\theta=45$ and can be ignored. Hence
\begin{align}\label{10.59}
\Theta=\frac{i}{\sqrt{2}}\left(exp(-\theta)-exp(-45)\right).
\end{align}
Similarly we can also have as a solution,
\begin{align}\label{10.59a}
\Theta=\frac{i}{\sqrt{2}}\left(exp(-\theta)-exp(-225)\right).
\end{align}
What we have here are the amplitudes of the gravitational waves (or electromagnetic waves) propagating along the radial vector at $45^\circ$ or $225^\circ$ to the z-axis. We can fix our coordinate system in such a way that the z-axis passes through the centers of the two objects of a binary system such as the GW150914 binary black holes. So the z-axis will automatically pass through the mass center of the binary objects. Then we make the perpendicular x-axis to pass through the mass center of the binary objects. So the radial vector will connect the mass center of the binary objects at the origin and a graviton or photon at the other end. The radial vector will make $\theta=45^\circ$ or $\theta=225^\circ$ corresponding to $\tan\theta=1$ solution. Similarly we can also have solution for $\tan\theta=-1$ which will be slightly modified version of the above solution. In this case the radial vector will make an angle of  $\theta=135^\circ$ or $\theta=315^\circ$ with the z-axis. The end result will be,
\begin{align}\label{10.59b}
\Theta=\frac{i}{\sqrt{2}}\left(exp(135/2)-exp(\theta/2)\right).
\end{align}
\begin{align}\label{10.59c}
\Theta=\frac{i}{\sqrt{2}}\left(exp(315/2)-exp(\theta/2)\right).
\end{align}
When $\theta=45$ in Eq.~\eqref{10.59}, the amplitude is zero. Small variation in $\theta=45\pm 0.05$ gives amplitude variation of $\Theta=\pm 1\times 10^{-21}$. These are the experimentally measured values of the gravitational-wave strain amplitude for GW150914 \cite{129}. GW amplitudes corresponding to Eq.~\eqref{10.59a} are of the order $10^{-98}$ and may be emitted by binary pulsars and therefore off-limit to existing GW detectors. GW amplitudes corresponding to Eq.~\eqref{10.59b} and Eq.~\eqref{10.59c} are of the order $10^{29}$ and $10^{68}$ respectively and may have existed in the early universe and therefore not accessible to us at the present epoch.\\
The coordinate set-up described above will spin about the y-axis with half the frequency of the gravitational waves which is equal to the orbital frequency of the binary objects $f/2$. As the binary objects approach the ring down frequency as in the case of black hole merger, the electromagnetic waves traveling along the same radial vector will also experience sharp increase in the frequency of light waves causing a gamma ray burst. This phenomenon was experimentally observed by Fermi Gamma-ray Burst Monitor in case of GW150914 \cite{159}. Present analysis provides the mechanism of this mysterious phenomenon for the first time. Due to the note following Eq.~\eqref{10.41} the angular position of the radial vector is restricted to a fixed value in the inertial frame of the coordinate system resulting in zero angular velocity which allows only one angular momentum eigen value $l=0$, giving $\lambda=0$. Substitution of $\lambda=0$ in Eq.~\eqref{10.50} does not affect the outcome in Eq.~\eqref{10.51} because $+\infty$ and $-\infty$ cancel out.\\
The radial vectors other than the four described above are associated with the non-zero angular momentum in the inertial frame  of the coordinate system. Hence the graviton or photon cannot break away from the gravitational pull of the central source and travel radially toward the observer on earth. The minimum radial distance at which the graviton or photon can embark on its radial journey is calculated in the next Section~\ref{rv}. It is obvious that the mechanism described above is equally applicable to millisecond binary pulsars and short GRBs, and could be the chief source of Galactic Centre Excess \cite{160}.

The formalism described above works only when angles are introduced in degrees. It does not work for radians because of the exponential terms. $\pm 0.05^\circ$ variation which gives the maximum GW150914 amplitude measured on earth is centered on the mass center of the combined BBH masses at source which is located at a luminosity distance of $D=410$ Mpc from earth. Thus the maximum amplitude at source can be expected to be $h\times D=1.265\times 10^4$ (unitless). The angle that this amplitude will envelope at the center of mass of BBH can be calculated from the same formula Eq.~\eqref{10.59}. It is $54.792^\circ$. Therefore amplitude variation of $\Theta=\pm 1.265\times 10^{4}$ at source is caused by angular variation of $45\pm 54.792^\circ$. Now it becomes possible to find the location of this maximum amplitude w.r.t. the center of mass of BBH. It can be given by
\begin{align}\label{10.59d}
r\mathbf{\hat n}=\frac{1.265\times 10^4}{tan(54.792^\circ)}\mathbf{\hat n}=8.927\times 10^3 \: \text{meters},
\end{align}
which is very much inside the event horizon at $1R_g=9.6005\times 10^4$. Here $\mathbf{\hat n}=1$\:meter is the unit vector along the radial vector $\mathbf{r}$. The agitation caused by gravitational waves causes the cosmic rays orbiting inside the event horizon to be pushed out of the event horizon causing $\gamma$-ray burst due to reversal of gravity assisted wave stacking which creates cosmic rays from the $\gamma$-rays as discussed in Section~\ref{Inside}.       

\subsubsection{Radial vector graviton spin and GRB}\label{rv}
We will rewrite the radial equation~\eqref{10.28} in dimensionless form by introducing a unitless independent variable $\rho$ so that 
\begin{align}\label{10.60}
&\rho=\left(\frac{\dot{\omega}r^3}{GM}\right)^2=\left(\frac{\dot{\omega}r^{5/2}}{GM}\right)^2r=\alpha r.\\
&\partial \rho=6\alpha\partial r.
\end{align}
Substitution of Eq.~\eqref{10.60} in Eq.~\eqref{10.28} gives
\begin{align}\label{10.61}
36\frac{\partial}{\partial \rho}\left(\rho^2\frac{\partial R}{\partial \rho}\right)+\rho R-\Gamma R=0.
\end{align}
Dividing Eq.~\eqref{10.61} by $\rho^2$ we get
\begin{align}\label{10.62}
\frac{36}{\rho^2}\frac{\partial}{\partial \rho}\left(\rho^2\frac{\partial R}{\partial \rho}\right)+ \left(\frac{1}{\rho}-\frac{\Gamma}{\rho^2}\right)R=0.
\end{align}
As far as the leading terms are concerned, for sufficiently large $\rho$ it is apparent that $R(\rho)=\rho^n e^{\pm\frac{1}{2}\rho}$ satisfies Eq.~\eqref{10.62} when $n$ has any finite value. This suggests that we look for an exact solution of Eq.~\eqref{10.62} of the form
\begin{align}\label{10.63}
R(\rho)=F(\rho)e^{-\frac{1}{2}\rho}
\end{align}
where $F(\rho)$ is a polynomial of finite order in $\rho$. Substitution of Eq.~\eqref{10.63} into Eq.~\eqref{10.62} gives equation for $F(\rho)$ as
\begin{align}\label{10.64}
F''+\left(\frac{2}{\rho}-1\right)F'+\biggl[\frac{1}{4}-\frac{1}{\rho}+\frac{1}{36\rho}-\frac{\Gamma}{36\rho^2}\biggr]F=0.
\end{align} 
\begin{align}\label{10.65}
F''+\left(\frac{2}{\rho}-1\right)F'+\biggl[\frac{1}{4}-\frac{35}{36\rho}-\frac{\Gamma}{36\rho^2}\biggr]F=0.
\end{align} 
Now we find a solution for $F$ in the form
\begin{align}
F(\rho)=\rho^s(a_0+a_1\rho+a_2\rho^2+\cdots)=\rho^sL(\rho),\quad
\ a_0\neq 0,\ s\geq 0.\label{10.66}
\end{align}
Substitution of Eq.~\eqref{10.66} into Eq.~\eqref{10.65} and dividing through by $\rho^{(s-2)}$ gives us the equation for $L$.
\begin{align}\label{10.67}
	\begin{split}
\rho^2L''+\rho(2(s+1)-\rho)L'+\biggl[\left(s^2+s-\frac{\Gamma}{36}\right)
-\left(\frac{35}{36}+s\right)\rho+\frac{1}{4}\rho^2\biggr]L=0.
	\end{split}
\end{align}
If we set $\rho=0$ in Eq.~\eqref{10.67}, it follows from Eq.~\eqref{10.66} that
\begin{align}\label{10.68}
s^2+s-\frac{\Gamma}{36}=0.
\end{align}
This quadratic equation in $s$ has two solutions,
\begin{align}\label{10.69}
s=-\frac{1}{2}\pm\frac{1}{6}(\Gamma+9)^\frac{1}{2}.
\end{align}
The boundary condition that $R(\rho)$ be finite at $\rho=0$ requires that we choose upper sign for $s$. With this, Eq.~\eqref{10.67} reduces to
\begin{align}\label{10.70}
\rho L''+\{2(s+1)-\rho\}L'+\left(\frac{1}{4}\rho-\frac{35}{36}-s\right)L=0.
\end{align}
Equation ~\eqref{10.70} can be solved by substituting Eq.~\eqref{10.66}. The recursion relation between the coefficients of successive terms of the series is observed to be
\begin{align}\label{10.71}
a_{\nu+1}=\frac{(35/36)+s-(\rho/4)}{\nu(\nu+1)+2(\nu+1)(s+1)-(\nu+1)\rho}\:a_\nu.
\end{align}
We can see from Eq.~\eqref{10.71} that the series can terminate when the following condition is satisfied.
\begin{align}\label{10.72}
\frac{1}{4}\rho-\frac{35}{36}-s=0.
\end{align}
Substituting $\rho$ from Eq.~\eqref{10.60} and $s$ from Eq.~\eqref{10.69} we get
\begin{align}\label{10.73}
\Gamma=\left\{\biggl[\frac{3}{2}\left(\frac{\dot{\omega}r^3}{GM}\right)^2-\frac{17}{6}\biggr]^2-9\right\}.
\end{align}
\begin{align}\label{10.74}
r=\left(\frac{2GM}{\dot{\omega}}\right)^{1/3}\biggl[\frac{17}{36}+\frac{1}{6}(\Gamma+9)^{1/2}\biggr]^{1/6}.
\end{align}
where from Eq.~\eqref{10.29} we have $\Gamma=j(j+1)$ \cite{77}. With respect to the four radial vectors restricted to a fixed angular position in the the inertial frame of the coordinate system, we have the angular momentum eigen value $l=0$. Hence the value of $\Gamma=j(j+1)$ in Eq.~\eqref{10.74} is exclusively decided by the spin of the graviton. The corresponding radial vector for $j=0, 1, 2$ are as given below. I have used following GW150914 data \cite{129} for computing radial distance $r_j$ for given $j$. These are the radial distances at which the series terminates for the given value of $j$. We have $f=150$ Hz., $m_1=36 M_\odot$, $m_2=29 M_\odot$, $M=(m_1+m_2)$, chirp mass $\mathcal{M}=28.09M_\odot$, $\dot{\nu}=\dot{f}=14344.907$. We get $r_{j=0}=5.7366\times10^5$m, $r_{j=1}=5.7873\times10^5$m and $r_{j=2}=5.8715\times10^5$m. These radii are greater than the photon capture radius $\sqrt{27}R_g$. They give the point of origin of gravitational wave which is spin dependent. If you can measure this radius, you know the spin of the graviton.\\
Fermi Gamma-ray Burst Monitor (GBM) had detected a 50 KeV weak hard X-ray transient lasting around 1s at the time of GW150914 event \cite{159}. This is equal to gamma-ray frequency  $\nu=1.21\times 10^{19}$ Hz. Arrival time was $0.4$s after the GW event. So the question is whether the short GRB photon can escape the black hole along the radial vector with the given energy and arrive at the point of departure of GW within $0.4$s. For this we will define radius $r_e$ of the event horizon using Eq.~\eqref{4.15a} which can be written as
\begin{align}\label{10.75}
r_e=\sqrt{\frac{\nu}{\dot{\nu}}\frac{GM}{c}}.
\end{align}
If we put
\begin{align}\label{10.76}
\frac{\nu}{\dot{\nu}}=\frac{GM}{c^3},
\end{align}
we get the $1R_g$ radius of the event horizon in PR. We do not have a theory to get the correct estimate of $\dot{\nu}$ using chirp mass for the frequency evolution of the short gamma-ray burst. Eq.~\eqref{4.14a} is applicable to gravitational waves only. However, we will use this equation here for short GRB to get some rough estimate. Using the chirp mass we get $\dot{\nu}=1.4057\times 10^{66}$. Photon spin is 1. For $l=0$ and $j=1$ we get $r_{j=1}=1.2553\times10^{-15}$m. This indicates that GRB photon originated at the mass center of the GW150914.  If we look at above equations we find that due to very high value of $\dot{\nu}$ the event horizon of the black hole has shrunk to $r_{j=1}=1.2553\times10^{-15}$m. GRB photons outside this radius are free to escape. This is possible only in PR, because in GR, $\dot{\nu}$ is zero for calculations involving event horizon. In PR light can accelerate at constant velocity $c$. This gives rise to the de Broglie force which can overcome the strong field gravitational force and shrink the event horizon. In GR conservation of energy is valid only locally and the invariance of linear space-time is global. In PR conservation of energy and invariance of periodic time (and wavelengths) are global, but linear time and linear distance are valid only locally. For light wave, $\lambda/T=c$ is a global constant and $d/t=c$ is a local constant. 

\section{Discussion and conclusion}
In Periodic Relativity (PR) we have proposed the singular motionless state of the primal energy called Fundamental substance of the universe which is devoid of any vibrations or motion and hence does not interact with any form of manifest energy. It is formless (having no boundary) and therefore infinite in extent. It is a unified field of all forms of energies (known and unknown to physics) and therefore indivisible. The Fundamental Substance replaces the space-time fabric of general relativity (GR). It has all the properties of empty space and cannot be detected by Interferometer because there are no waves or oscillations or motion in it. The entire universe exists within it, and it permeates everything in the universe. This motionless Fundamental Substance is not the energy but becomes energy when a small portion of it begins to move and apparently divide into particle waves having wavelengths and periods. This is how motion, energy, wavelengths (space) and periods (time) are simultaneously created and the infinite becomes limited with various forms and the Absolute One becomes many; thus the laws of relativity becomes operational. All other relativistic discussions in PR are in support of this primary prediction.

It is well known that quantum mechanics (QM) and general relativity (GR) have remained two incoherent theories for more than a century. All the fundamental derivations of quantum mechanics originate from the energy momentum invariant defined by Eq.~\eqref{16.1} and all the fundamental derivations of general relativity originate from the line element defined by Eq.~\eqref{16.2}. 
\begin{align}\label{16.1}
E^2=(m_0c^2)^2+(\mathbf{p}c)^2.
\end{align}
\begin{align}\label{16.2}
ds^2=c^2(1-r_s/r)dt^2-(1-r_s/r)^{-1}dr^2-r^2d\theta^2-r^2\sin^2{\theta}d\phi^2,
\end{align}
where $r_s=2GM/c^2$. One cannot derive Eq.~\eqref{16.2} from Eq.~\eqref{16.1} and vice versa. So all the coherence within QM is founded upon Eq.~\eqref{16.1} and all the coherence within GR is founded upon Eq.~\eqref{16.2}. And they don't talk with each other. As a consequence GR does not recognize the global invariance of energy momentum equation~\eqref{16.1} and QM has nothing to do with the line element of GR Eq.~\eqref{16.2}. 

In periodic relativity (PR) we modified both these equations to make them compatible with each other, at the same time satisfying Einstein's field equations \cite{44,60,153}. This is done by introducing a deviation factor $n$ in both the equations as follows. As a consequence we can now derive Eq.~\eqref{16.4} from Eq.~\eqref{16.3} and vice versa.     
\begin{align}\label{16.3}
E^2=(m_0c^2)^2+n(\mathbf{p}c)^2.
\end{align}
\begin{align}\label{16.4}
ds^2=c^2dt^2-n(dr^2+r^2d\theta^2+r^2\sin^2{\theta}d\phi^2).
\end{align}
Line element Eq.~\eqref{16.4} satisfies Einstein's field equations when following conditions given by Eq.~\eqref{4.58a} are satisfied \cite{44}.
\begin{align}\label{4.58aa}
\left(\frac{r}{n}\frac{\partial n}{\partial r}\right)=0 \qquad and \qquad \left(\frac{r}{n}\frac{\partial n}{\partial r}\right)=-4.
\end{align}
The first condition is satisfied when deviation factor $n$ is any real number constant. Hence, now we can seek coherence within PR in energy equation given by Eq.~\eqref{16.3} \cite{153}. This was not possible in GR which is a geometrical theory. The invariance of the modified energy momentum (MEM) equation~\eqref{16.3} is universal. An array of different topics covered in PR convey the idea that everything in the universe has its origin in energy and it is not just a mathematical parameter for converting one form of energy into another. Einstein's mass energy equivalence was a big step in the world of physics toward this realization. We have presented derivation for the deflection of light around the sun from fundamentals by introducing parameter $\psi$ ignored by general relativity and by introducing vectors. Deviation factor $n=1$ for light. Factor $\left(\cos{\psi}+\sin{\psi}\right)$ causes the geodesic like trajectory of the particle. Here we can relate the additional component of acceleration with the rotation of the velocity vector which causes the curvature of the trajectory. We have distinguished the Cartesian curvilinear acceleration from the polar conic acceleration and explained why they are not equal even though they are derived from the same velocity vector. We have derived expression for the Lorentz invariant acceleration. 

PR proposes that all massive fundamental particles of physics can disintegrate into massless particles traveling at speed of light as they approach the Planck energy level. This is how all fundamental particle fields are united into a single field associated with the particle described as Saviton (principle quantum number $n=1$) in article \cite{154}. In PR we are at liberty to choose different deviation factor $n$ for different two body systems so the coherence is sought in Einstein's field equations and not in the line element as it is done in GR by applying the same weak field approximation to all kinds of two body systems. This works well within solar system but it has failed to predict the correct radius of the event horizon of a black hole which is $1R_g$ and not $2R_g$. PR explains $\gamma$-ray production in M87 BH due to  gravity assisted EM wave stacking and $\gamma$-ray burst due to reversal of the same process. GR cannot explain these phenomena. Similarly GR has nothing to say about shrinking of the event horizon due to the short Gamma-ray burst. So coherence isn't helping GR in this regard.

GR has locked itself into using the manipulated factor 2 in the weak field approximation which also appear in the Schwarzschild radius for the reason that GR does not have a mechanism to explain the relation between the coordinate time and the proper time of a body in motion. It is just that the three major tests of GR within solar system fits well with factor 2 in the weak field approximation. In PR we have the mechanism to explain the relation between the coordinate time and the proper time of a body in motion. The deviation factor $n$ in PR may have an internal structure dependent on the natural frequency and composition of the orbiting body (Doppler frequency shift of the constituent massive particles of the body in motion which decides the value of $n$ which can be different for different two body systems). Therefore it is proper to seek coherence between array of different topics covered in PR in the MEM invariant Eq.~\eqref{16.3}. Unlike GR, the gravitational redshift Eq.~\eqref{4.3} and gravitational wave equation~\eqref{4.18a} are both derived using MEM invariant Eq.~\eqref{16.3}.  

In periodic quantum gravity and cosmology theory \cite{154}, we have proposed a wave equation given by 
\begin{align}\label{b23}
 \begin{split}
i^2\hbar^2\frac{\partial^2 \psi}{\partial t^2} &=\left(1-\frac{(Hr)^2}{c^2}\right)^{1/2}\frac{G\eta H}{r}\left[\frac{-\hbar^2}{c^2}\boldsymbol{\nabla}_j^2+m_o^2\right]\psi\\
&+\frac{4}{3}\frac{\alpha_s \hbar cH}{r}\psi+\frac{\sigma Hr}{\hbar c}\psi+\frac{Zke^2H}{r}\psi.
 \end{split}
\end{align}
Here $G$ is the gravitational constant and $H$ the Hubble parameter. Solution of this wave equation can generate the entire table of fundamental particles of the standard model by selecting different values of principle quantum number $n$. In this wave equation we have not utilized the deviation factor $n$ of periodic relativity which when used 
replaces the potential energy parameter $V$ as discussed in our article on Hydrogen spectra \cite{153}. So both cannot be used simultaneously. Generally in quantum gravity theories, $H$ is assigned an arbitrary high value prior to symmetry breaking at Planck epoch and as $H$ rolls down to the low value at present epoch all the energy is gradually released into our universe. No body knows where it comes from. Zero energy universe theory requires existence of negative gravitational energy to balance the positive energy of the universe but there is no such evidence for its existence. So in our theory \cite{154}, the initial state of the universe is described by energy $E=0$ and Hamiltonian $H=0$, because we have said that the energy comes from the motionless fundamental substance of the universe when it begins to move. Only difference between the infinite Fundamental Substance and the finite energy, is in motion. Like Ocean and its waves. The content is the same. The content never increase or decrease, no matter how much energy come into the universe. Another success of our theory is that we have solved the problem of time encountered in quantum gravity theories by associating Hubble parameter with Planck frequency at Planck time. Hubble parameter has units of frequency and the inverse of Hubble constant at the present epoch gives the age of the universe. So this is another application of periodic time. It locks the maximum value of Hubble parameter at Planck time. In the neighborhood of Planck epoch all the free moving fundamental particles of physics obey the periodic invariant proposed in article \cite{44} and in Eq.~\eqref{1.2} which can be written as
\begin{align}\label{16.5}
s^2=\lambda^2-V^2T^2,
\end{align}
where $\lambda = h/p$ \ is the associated de Broglie wave\-length, $V = c^2/v$ the phase velocity, $v$ the particle velocity, and $T$ the period of the wave. Here linear space and linear time are eliminated and therefore deviation factor $n$ is not required because wavelength and period are naturally connected as continuum. Linear time and linear distance of relativistic invariant are not naturally connected, hence you need the deviation factor. For particles like photons and savitons \cite{154}, periodic invariant reduces to what we call quantum invariant Eq.~\eqref{2.3} which is described by only two parameters. Energy and frequency. Reminder here that saviton (principle quantum number $n=1$) in article \cite{154} gives birth to all the fundamental particles of standard model (principle quantum numbers $n>1$) including photon plus many particles unknown to physics. 
\begin{align}\label{16.6}
s^2 = (h/E)^2 - (1/\nu)^2.					
\end{align}
During a big crunch, quantum invariant can vanish in an absolute sense when $E > E_p$ and 
$\nu > \nu_p$, where $E_p$ and $\nu_p$ are Planck energy and Planck frequency respectively. In this case when the particle tries to acquire more energy than the Planck energy, it will violate the Compton limit on the wavelength and thus the particle wave will collapse and will become perfectly motionless leaving no mass gap. This mechanism is described in Section~\ref{gemw} following Eq.~\eqref{b66b}. Thus the space-time continuum connecting two points gets completely obliterated and the resulting sub-quantic medium resembles a singularity characterised by a perfectly motionless indivisible field which is infinite like empty space. This is the fundamental substance of the universe from which a specific finite excitation had arisen as a wave particle duality. Such a singularity suggests a motionless field devoid of ripples capable of giving birth to energy which is always in motion. This describes the ultimate coherence of all forms of energy in the universe. Eq.~\eqref{1.26j} (with plus sign only) also suggests the particle wave collapse for $x=z=0$. Whether it happens at Planck energy level or some other energy level, is a question for debate. 

We were able to predict the limiting radius of the event horizon of a spinning black hole to be $1R_g$. We could explain the shadow of a black hole by introducing de Broglie acceleration in the balance equation of gravitational and the centrifugal accelerations. The resulting formula for the photon ring diameter is independent of the mass or the spin of the black hole. The formula shows that the shadow of a black hole is due to transition of light frequency into ultraviolet and higher range. PR does not require a special line element for a spinning black hole.

In periodic quantum gravity and cosmology theory \cite{154} the universe begins with a vibration in the fundamental substance of the universe which is singular motionless, indivisible and infinite. This is how motion, energy, space (wavelength), and time (period) are simultaneously created. With these energy quantas, gravity also makes its appearance. When the waves subside, gravity also disappear. Wave particle duality, entanglement and gravity are due to indivisibility of this fundamental substance. This is the primary prediction of PR. Theory of gravity in flat space time is required to unify the quantum physics with periodic relativity. This allowed us to introduce Einstein's field equations in hydrogen atom model and replace the potential energy term of Schrodinger equation with the deviation factor $n$ as a ratio of (Coulomb) accelerations. Rotation curves of galaxies uses the same approach where deviation factor is introduced as a ratio of (Newtonian) accelerations \cite{60}. This formalism then makes it possible to predict the Lamb shift \cite{153} and quantify perihelic precession of electron in Hydrogen atom. 

We have shown that we can introduce gravity into electromagnetic wave formalism by working with wavelengths and periods of the wave for eliminating linear distance and linear time. Without this basic approach it is not possible to introduce gravity within electromagnetic wave. This is impossible in general relativity. In case of light waves, two successive photon quantas can fall toward or away from each other due to gravity with accelerated motion while the light travels at a constant velocity c. This is the mechanism of gravitational frequency shift when photons acquire or loss energy. Concept of gravity assisted wave stacking is introduced for acquiring and emitting energy which can produce $\gamma$-rays from synchrotron hard x-rays in black holes, or $\gamma$-ray bursts from cosmic rays crossing the event horizon of a black hole. Same theory apply to Savitons at Planck epoch \cite{154}. Working with wavelengths and periods is a natural way of quantizing space and time. But this require that we first develop a theory of gravity in flat space time, which we already did. We showed that the electromagnetic wave is held together by gravitational forces just as a planet or solar system is held together by gravity. We also showed how linear time in Eq.~\eqref{1.22c} can be replaced by periodic time. This provides mathematical proof for the periodic nature of time and existence of very powerful gravitational radiation at Planck epoch. All this is possible due to the introduction of relativistic mass in the formalism. Theory points out the existence of de Broglie force which gives equation of propagation of massless bosons. This invariant equation yields the mass of graviton from GW150914 data to be $m_g<1.51\times 10^{-41}$eV/$c^2$. Same equation yields gravitational redshift of gravitational waves which is greater than that predicted in GR. Ratio of Newtonian acceleration and de Broglie acceleration is utilized in quantum mechanical formalism for deriving formula for gravitational-wave strain amplitude. A method of determining graviton spin is discussed. Mechanism that explains Gamma-ray burst associated with GW150914 is presented. Same can explain millisecond pulsars and Galactic Centre Excess. It is shown that a very high rate of light frequency of GRB can shrink the event horizon of a black hole. This theory provides reasonably accurate way of introducing gravity in quantum mechanics.

\section{Acknowledgment}
Author is grateful to several experts in the field for comments and suggestions. 

\pdfbookmark{References}{Ref}
\bibliographystyle{amsalpha}

\end{document}